\documentclass[prb,twocolumn,showpacs,preprintnumbers,amsmath,amssymb]{revtex4-1}

\usepackage{graphicx}
\usepackage{dcolumn}
\usepackage{bm}
\usepackage{amsmath}
\usepackage[dvips]{color}

\begin{document}

\title{
Interplay between Kondo and Andreev-Josephson effects 
 in a quantum dot coupled to one normal and 
two superconducting leads 
}

\author{
Akira Oguri,$^1$
Yoichi Tanaka$^2$ and 
Johannes Bauer$^{3,4}$ 
}

\affiliation{
$^{1}$Department of Physics, Osaka City University, Osaka 558-8585, 
Japan
\\
$^{2}$Condensed Matter Theory Laboratory, RIKEN, 
Saitama 351-0198, Japan
\\
$^{3}$
Max-Planck Institute for Solid State Research, 
 Heisenbergstr.1,
 70569 Stuttgart, Germany\\
$^4$
Department of Physics, Harvard University, Cambridge, Massachusetts 02138, USA
}

\date{\today}

\begin{abstract}
We study low-energy transport through a quantum dot coupled to one normal 
and two superconducting (SC) leads in a junction of Y-shape.  
In this geometry a crossover between Kondo dominated and Cooper-pairing
dominated states occurs by tuning the parameters such as  the quantized energy 
level  $\epsilon_d$  of the dot and  
the Josephson phase $\phi$, which induces a supercurrent flowing 
between the two SC leads through  the dot.  
Furthermore,  Andreev scattering takes place at the interface between the dot
and normal lead. The low-lying energy states of this system can be described  
by a local Fermi-liquid theory for interacting Bogoliubov particles. 
In a description based on an Anderson impurity model we calculate transport 
coefficients, renormalized parameters and spectral function, 
using Wilson's numerical renormalization group (NRG) approach, 
in the limit of large SC gap. 
Our results demonstrate how the Andreev resonance level approaches 
the Fermi level in the crossover region between Cooper-pairing singlet state 
and strong coupling situation as $\epsilon_d$ or $\phi$ are varied. 
The strong coupling situation shows a Kondo effect with a significantly 
renormalized resonance width. 
The crossover is smeared when the coupling between the dot and  normal 
lead is large.  
Furthermore, asymmetry in the Josephson junction 
suppresses the cancellations of the SC proximity for finite $\phi$, 
and it favors the SC singlet state rather than the Kondo singlet.
\end{abstract}


\maketitle

\section{Introduction}

The Kondo effect in superconducting (SC) materials 
has been one of the major topics 
in condensed matter physics over forty years.
The energy gap $\Delta_\mathrm{SC}$ of a superconductor suppresses 
the  magnetic screening of the conduction-electrons at low 
temperatures below the Kondo temperature $T_K$.  
The competition of these 
effects causes a quantum phase transition
(QPT) between a magnetic-doublet and  nonmagnetic-singlet ground states,  
which emerge at $\Delta_\mathrm{SC} \gg T_K$ 
and $\Delta_\mathrm{SC} \ll T_K$, respectively. 
\cite{ShibaSoda,MullerHartmanZittartz,Matsuura,JarrelSiviaPatton,SSSS,YoshiokaOhashi,sc3,Florens,Bauer2012}

The QPT has also been studied intensively 
for quantum dots,\cite{MartinYeyati} 
and experiments have been carried out 
for carbon nanotube and semiconductor quantum dots.
\cite{Buit:2002,Eich:2007,Sand:2007,Buiz:2007,Grov:2007}
One of the merits of quantum dots is the high tunability, 
and various types of the configurations can be examined.
For instance, 
for a quantum dot (QD)  embedded between 
 two superconducting leads in a SC/QD/SC configuration,
the competition between the Kondo and Josephson effects 
 has also been expected to occur.
\cite{Ishizaka,ClerkAmbegaokar,RozhkovArovas,Avishiai,Vecino,siano_egger,Kusakabe,Choi,sc1,sc2,sc4,Hecht,CT_QMC}
 Furthermore,
an interplay between Andreev scattering and the Kondo effect 
has been studied experimentally\cite{Graber,Deacon}
and theoretically\cite{Fazio,Schwab,Cho,Clerk,Sun,Cuevas,Avishai1,Aono,Krawiec,Splett,Doma,TanakaNQDS,TanakaNTDQDS,TanakaNTDQDS2} for a QD connected to  
 a normal-metal (N) lead and  a SC lead in a SC/QD/N configuration.

\begin{figure}[b]

\leavevmode
\includegraphics[width=0.85\linewidth]{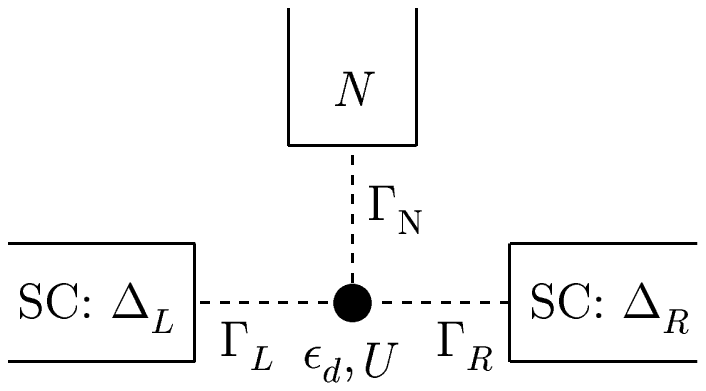}

\caption{
Anderson impurity ({\large $\bullet$}) 
coupled to one normal and two superconducting leads: 
 $\epsilon_d$ and $U$ are the level position and Coulomb interaction.
$\Gamma_{\nu} \equiv \pi \rho \,v^{2}_{\nu}$   
with $\rho$ the density of states of the leads, and
$v_{\nu}$  the tunneling matrix element ($\nu=L,\,R,\,N$).
The complex SC gap $\Delta_{L/R} =|\Delta_{L/R}|\, e^{i\theta_{L/R}}$ 
causes the Josephson current for finite 
$\phi \equiv \theta_{R}^{}-\theta_{L}^{}$.  
}
%
\label{fig:system}

\end{figure}


An interesting extension, which is also relevant
  experimentally,\cite{Y_experiment_Schoenenberger,Y_experiment_Herrmann,Y_experiment_Marcs} is a junction of Y-shape, 
at which a single QD is coupled to one normal 
and two superconducting (SC) leads as shown in Fig.\ \ref{fig:system}.
This system has been studied by  Pala,  Governale and  K\"{o}nig, 
using a real-time diagrammatic approach based on a perturbation expansion 
with respect to the tunneling matrix elements.
\cite{PalaGoveKonig_NJP,GovePalaKonig_PRB}  
Their calculations reveal precise features 
of the Josephson current and Andreev bound states 
both  in equilibrium and nonequilibrium situations   
where a finite bias voltage is applied between the dot and the normal lead.
However, their approach is  not applicable at low temperatures $T<T_K$, 
and thus the competition between superconductivity and the Kondo effect 
occurring in this system has been left to be explored.
Specifically, in the Y-junction the conduction electrons 
from the normal lead can screen the local moment induced 
in the QD. This changes 
the sharp transition between the magnetic and non-magnetic ground states   
into a crossover between a Kondo singlet and a Cooper-pairing singlet. 
Furthermore, the Andreev scattering, which takes place 
between the QD and normal lead, can be controlled through 
the phase difference $\phi$ between the order parameters of
two SC leads. This is because the SC proximity on the QD 
depends sensitively on the properties of the junctions, 
and thus on $\phi$. 
Conversely, the Josephson current flowing between the 
two SC leads is affected by the Andreev scattering 
of the conduction electrons from the normal lead.

The purpose of the present work is to study 
these interplays of the Kondo, Andreev, and Josephson effects 
which can be observed for the QD embedded  
in this three terminal system.\cite{sces2011}
To this end, we explore a wide region of the parameter space 
of this Y-junction, 
varying the position of a quantized energy level $\epsilon_d$ of the QD 
modeled with an Anderson impurity, 
and also examine how an asymmetry of the Josephson junction 
affects the low-temperature properties.
Specifically, we focus on the crossover between the ground states    
which can be classified into a Kondo singlet 
and a local Cooper-pairing singlet 
according to the fixed points of 
Wilson's numerical renormalization group (NRG).\cite{KWW2,Hewson1}

For strongly correlated systems the Coulomb interaction $U$ is 
larger than the other energy scales,
and for such cases the critical behavior near the QPT is scaled  
by a single parameter $\Delta_\mathrm{SC}/T_K$.   
In systems with QDs,  however, 
some of the parameters can be tuned experimentally, 
and $U$ is not always the largest energy scale.
Therefore, the ground-state properties depend on the other 
parameters, such as $\epsilon_d$, $U$, and the hybridizations 
between the dot and leads.
Specifically, for small interactions $U < \Delta_\mathrm{SC}$, 
the SC pair correlations can penetrate into the QD 
and create a local Cooper pair, consisting of a linear combination of 
an empty state and a doubly occupied state.    
The essential physics of the local Cooper pairing  
can be deduced from the fixed point of the NRG  
in the limit of $\Delta_\mathrm{SC} \to \infty$,
where the coherence becomes of the order of the lattice constant.
We consider in detail this large SC gap limit in the present work.


The low-lying energy states of this Y-junction 
can  be described by a local Fermi liquid of 
the interacting Bogoliubov particles.  
This is because the normal lead, coupled to the QD,  
has a continuous energy spectrum at the Fermi level. 
Furthermore, the local SC pair potential $\Delta_d^{}$ 
is induced in the QD by the proximity effect. 
This $\Delta_d^{}$ also plays a central role, 
and it varies sensitively with $\epsilon_d$ and $\phi$.
For instance, the conductance due to Andreev scattering   
can be expressed  at $T=0$   
in terms of  phase shift $\delta$ 
for the renormalized quasiparticles and 
the angle $\Theta_B$ of the Bogoliubov rotation 
determined by $\Delta_d$, 
and is enhanced at the crossover region 
between the Kondo singlet 
and local-Cooper 
pairing singlet.\cite{TanakaNQDS,TanakaNTDQDS,TanakaNTDQDS2}
We also calculate the renormalized parameters 
for the interacting Bogoliubov particles.
The results of the renormalization factor $z$, the Wilson ratio $R$, 
and the renormalized  Andreev level $\widetilde{E}_A$ 
that corresponds to quasiparticle peak position appearing 
in the spectral function of the QD, 
provide us with sufficient information to understand  
the ground-state properties of the system thoroughly.

This paper is organized as follows. 
In Sec.\ \ref{sec:model},   
we introduce the single impurity Anderson model for the Y-junction, 
and provide some examples which capture typical behavior near the QPT 
occurring in a SC/QD/SC junction with a finite SC gap. 
In  Sec.\ \ref{sec:large_sc_gap}, 
we give a local-Fermi liquid description 
 for the interacting Bogoliubov particles  
in the large SC gap limit, and present  
the expressions of the correlation functions in terms of 
the renormalized parameters. 
Then in Sec.\ \ref{sec:result}, 
we provide the NRG results for the spectral function, 
transport coefficient, and renormalized parameters.  
A summary is given in Sec.\ \ref{sec:Summary}.

\section{Model and fundamental aspects} 
\label{sec:model}

\subsection{Model}
We start with the Anderson impurity model for a single quantum dot coupled 
to one normal ($N$) and two superconducting (SC) leads, 
\begin{eqnarray}
H &=& H_{d} 
\, + \sum_{\nu =N,L,R} H_{\nu}  
+ \sum_{\nu =N,L,R} H_{T,\nu}  
\, +\, H_\mathrm{SC}^{} 
\:.
\label{Hami_seri}
\end{eqnarray}
The explicit form of each part is given by, 
\begin{align}
& H_{d}
=
\xi_{d} \left(n_{d}-1\right)
+\frac{U}{2}\left(n_{d}-1\right)^2
\nonumber\\
& 
H_\nu = 
\sum_{k,\sigma} \, \epsilon _{k}^{}
c_{\nu,k\sigma}^\dag c_{\nu,k\sigma}^{},
\nonumber \\
&
  H_{T,\nu} = \sum_{\sigma}  v_{\nu}^{}
 \left( \psi_{\nu,\sigma}^\dag d_{\sigma}^{} + 
  d_{\sigma}^{\dag} \psi_{\nu,\sigma}^{} \right) \;,
\nonumber\\
&H_\mathrm{SC}^{} \, 
=\, 
\sum_{k}\left(\Delta_{L}\, 
c_{L,k\uparrow}^\dag \,c_{L,-k\downarrow}^\dag 
+ \textrm{H.c.}\right)  \nonumber \\ 
& \qquad\quad  
+ \sum_{k}\left(\Delta_{R}\, 
c_{R,k\uparrow}^\dag \,c_{R,-k\downarrow}^\dag 
+ \textrm{H.c.}\right).
\label{Hami_seri_part}
\end{align}
Here, $\xi_{d}\equiv\epsilon_{d}+U/2$, and $U$  is the Coulomb interaction. 
The operator $d^{\dag}_{\sigma}$ creates an electron with 
energy $\epsilon _{d}$ and spin $\sigma$ at the dot, 
and $n_{d}=\sum_{\sigma}d^{\dag}_{\sigma}d^{}_{\sigma}$.
The operator $c_{\nu,k\sigma}^\dag$ creates an electron 
with the energy $\epsilon _{k}$ in 
the leads $\nu$ ($=N,L,R$).
The couplings between the dot and leads are described by 
the tunneling matrix elements $v_{\nu}^{}$, 
and a linear combination of the conduction 
$\psi_{\nu,\sigma}^{} \equiv \sum_k c_{\nu,k\sigma}^{}/N_\nu$ 
with $N_\nu$ the number of the states in each lead.
We assume that the density of states 
$\rho(\epsilon) \equiv \sum_k \delta(\epsilon-\epsilon_{k})/N_\nu$ 
and $\Gamma_{\nu}(\epsilon) \equiv \pi  v_{\nu}^{2}\,\rho(\epsilon)$ 
are constants independent of the frequency $\epsilon$ at $|\epsilon|<D$, 
where $D$ is the half band-width for the leads. 
The complex $s$-wave BCS gap, 
 $\Delta_{L/R}^{} = |\Delta_{L/R}^{}| e^{i\theta_{L/R}}$
for the SC leads on the left ($L$) and right ($R$) 
induces a Josephson current when the 
phase difference $\phi \equiv \theta_{R}^{} -\theta_{L}^{}$ is finite.   
In this three terminal system, 
the current  $J_{\nu}$ flowing from the dot to the lead $\nu$ 
is given by
\begin{align}
&J_{\nu} \,=\, \frac{ie}{\hbar}
\sum_{\sigma} v_{\nu}^{} 
\left( 
 \psi_{\nu,\sigma}^{\dag} d_{\sigma}^{}  
-
d_{\sigma}^{\dag} \psi_{\nu,\sigma}^{} 
\right).
\label{current}
\end{align}
Here, $-e$ denotes the electron charge with $e>0$.

The Hamiltonian $H$ contains a number of parameter regimes to be explored.  
We mainly consider the case where 
the couplings and the amplitude of the SC gaps are symmetric:  
 $\Gamma_L^{} = \Gamma_R^{}$ $(\equiv \Gamma_S/2)$ and  
$|\Delta_L^{}| = |\Delta_R^{}|$ $(\equiv \Delta)$, 
for simplicity.  
The asymmetry in the Josephson coupling   
 $\Gamma_L^{}  \neq \Gamma_R^{}$ is also examined 
in the last part in Sec.\ \ref{subsec:asymmetric_junction}.

\begin{figure}[t] 
 \leavevmode
\includegraphics[width=0.8\linewidth]{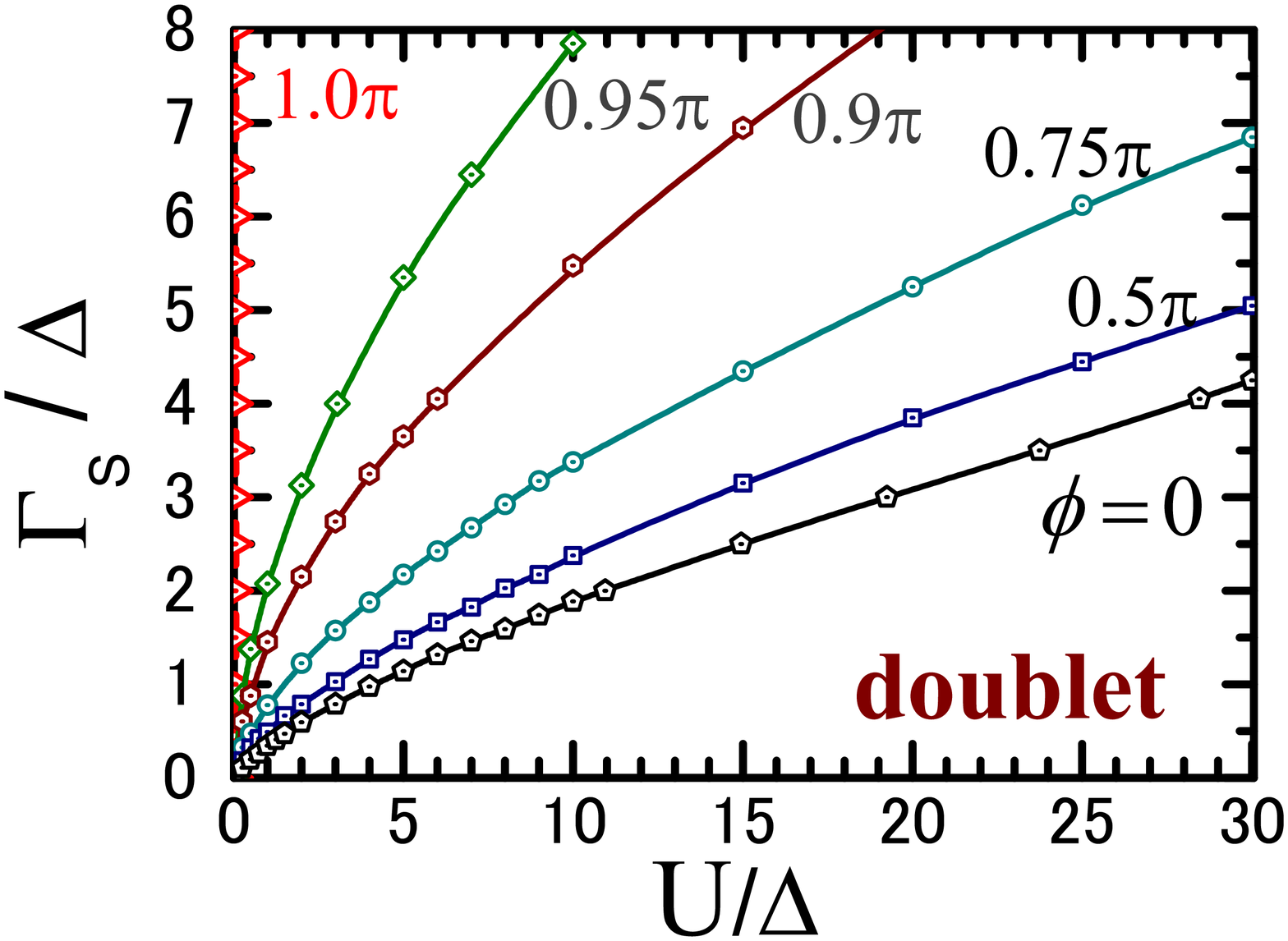}
\caption{
(Color online) 
NRG results for the ground-state phase diagram of  
the Anderson impurity connected to two SC leads 
for several value 
of 
 $\phi \equiv \theta_{R}^{} -\theta_{L}^{}$ 
in the electron-hole symmetric case $\epsilon_d = -U/2$.   
In the upper and lower sides of each boundary  
the ground state is 
a nonmagnetic singlet and magnetic doublet,
respectively.
 The Josephson junction is assumed to be symmetric 
 $\Gamma_L^{}=\Gamma_R^{}$ $(\equiv \Gamma_S/2)$ and 
$|\Delta_L^{}| = |\Delta_R^{}|$ $(\equiv \Delta)$.   
The normal lead is {\it not} connected $\Gamma_N^{}=0$ in this case. 
}
\label{Fig:results_2SC_GS}
\end{figure}

\subsection{QD connected two SC leads ($\Gamma_N^{}=0$)}

Before discussing the three terminal case,
we first of all consider a simpler case with $\Gamma_N=0$,  
where the normal lead is disconnected 
and the QD is coupled only to the two SC leads,  
in order to review  some typical features of 
the competition between the Kondo and Josephson 
effects.\cite{siano_egger,Choi,sc2,sc4,Hecht,Florens} 
In this case the QPT occurs as a level crossing of 
the lowest two energy states of $H$, 
and thus the expectation value for the Josephson current 
and that for the order correlation functions   
show a discontinuous jump at the critical point.

In Fig.\ \ref{Fig:results_2SC_GS}, 
the NRG results for the phase diagram of the ground state  
in the electron-hole symmetric case $\epsilon_d = -U/2$ 
is plotted in a $U/\Delta$ vs $\Gamma_S/\Delta$ plane 
for several values of $\phi$. 
The upper (lower) side of each boundary  
correspond to the parameter region where 
the ground state is a non-magnetic singlet (magnetic doublet).
These results clearly show that the magnetic-doublet region, 
appearing for large $U/\Delta$ or small $\Gamma_S/\Delta$, 
expands  as $\phi$ increases. 
Therefore, the phase difference $\phi$ between the two SC order parameters 
tends to suppresses the Kondo screening,\cite{siano_egger,Choi,sc2,sc4}
which in this case is carried out also 
by the quasiparticle excitations 
above the SC energy gap. 
As the value of $\phi$ increases from $0$ to $\pi$, 
the SC proximity effect penetrating from the left lead 
and that from right lead cancel each other at the impurity site. 
This is because the SC proximity  
is determined by a superposition of  $\Delta_L$ and  $\Delta_R$,   
which can be explicitly seen in the impurity Green's function 
given in Eqs.\ \eqref{eq:Gdd_def} and  \eqref{eq:G_lead_2}.
This suppression of the proximity effect leads to a reduction of the parameter region 
for singlet formation as seen in Fig.\ \ref{Fig:results_2SC_GS}. 
Specifically, at $\phi = \pi$ 
the system is equivalent to an Anderson impurity model 
with an insulating bath,\cite{sc2,GalpinLogan} 
where for the particle-hole symmetric case and $U>0$ the ground state 
is always a doublet because the impurity level situates just 
on the Fermi level in the middle of the band gap.

\begin{figure}[t] 
 \leavevmode
\begin{minipage}{1\linewidth}
\includegraphics[width=0.48\linewidth]{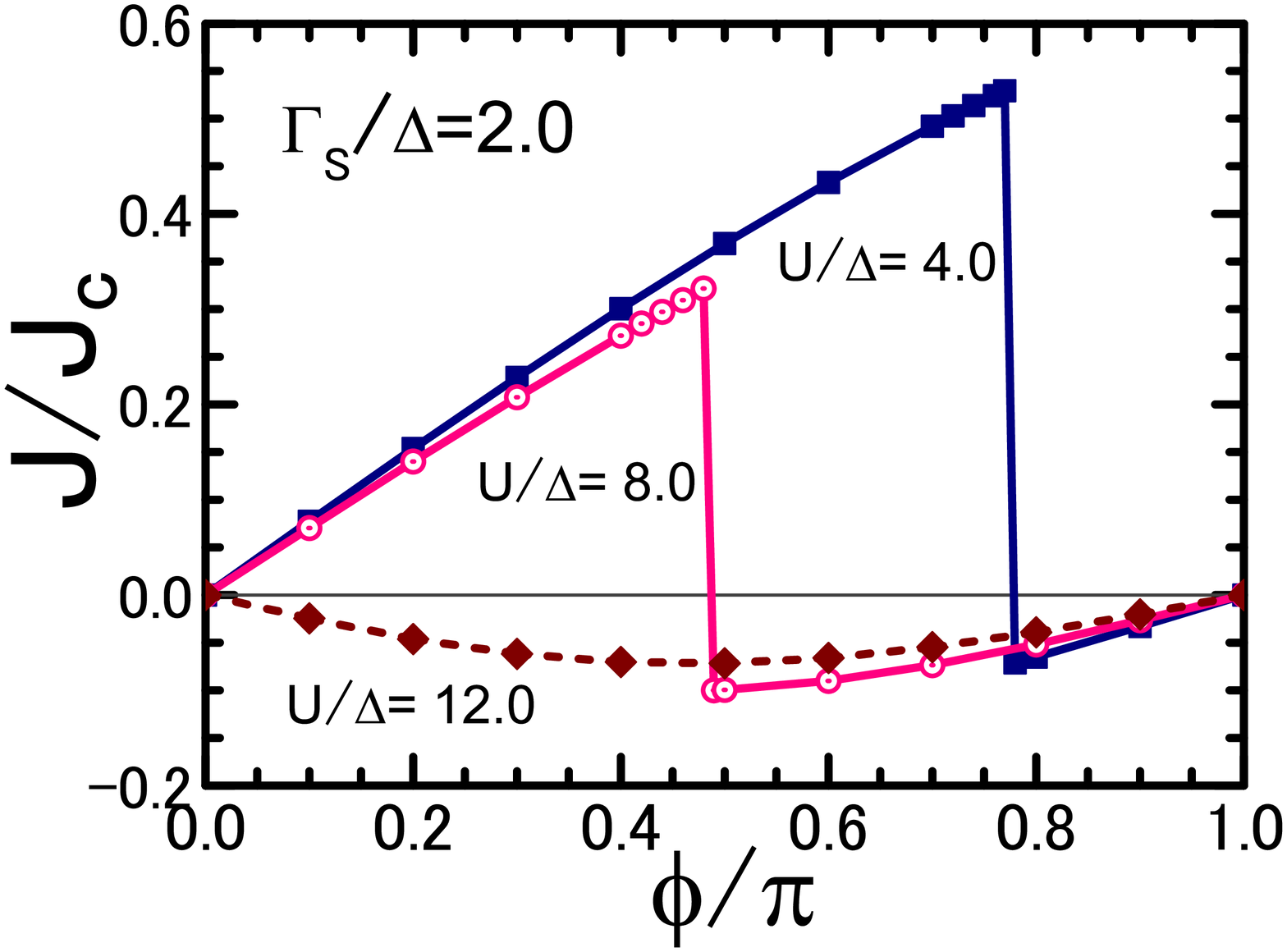}
\rule{0.01\linewidth}{0cm}
\includegraphics[width=0.485\linewidth]{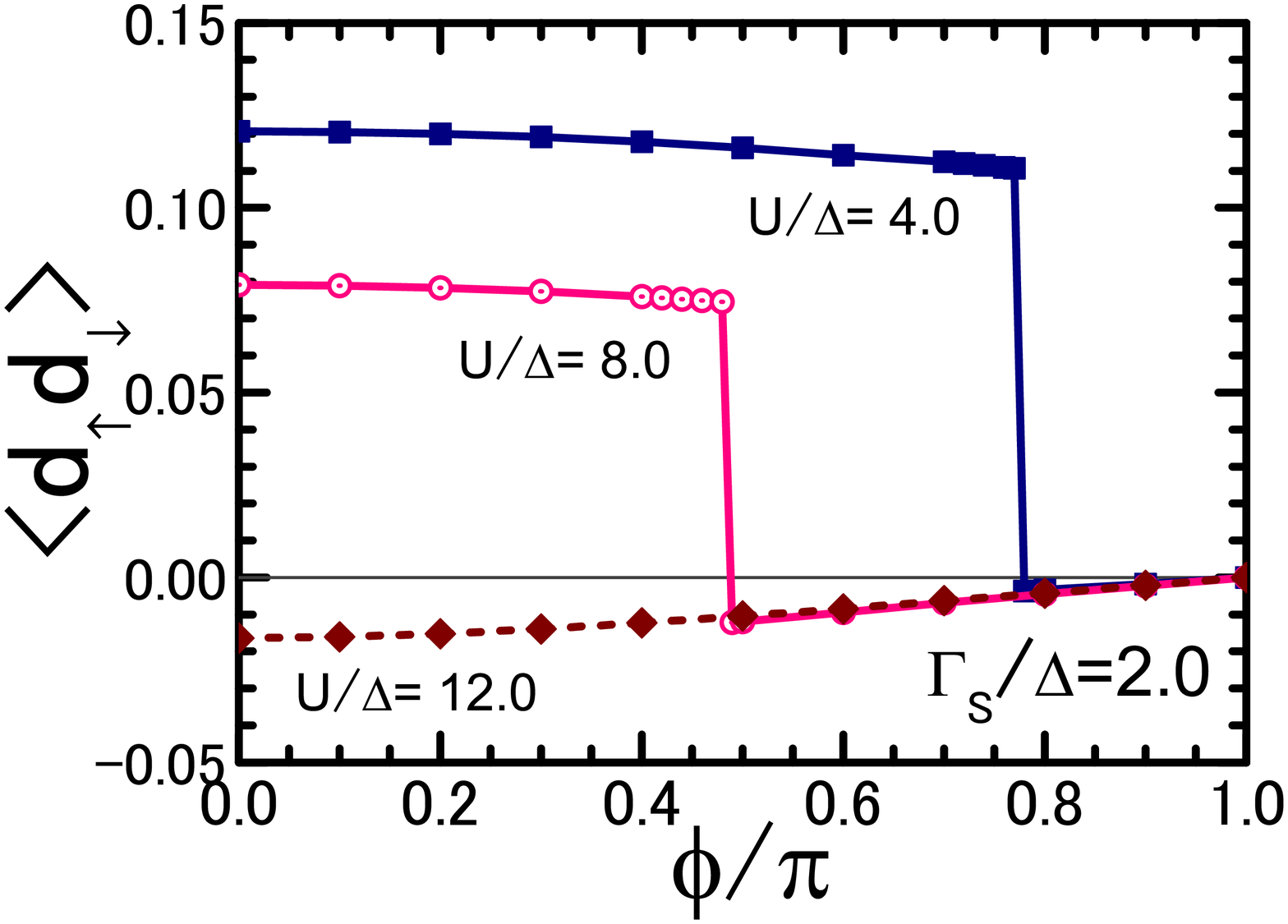}
\end{minipage}
\caption{
(Color online)
NRG results for the Josephson current and 
$\langle d_\uparrow^{} d_\downarrow^{}  \rangle$ 
for the Anderson impurity connected to two SC leads 
are plotted as functions of $\phi$ 
 for several value of $U$ 
in the electron-hole symmetric case $\epsilon_d = -U/2$. 
The other parameters are chosen such that  
 $\Gamma_L^{} = \Gamma_R^{}$ $(\equiv \Gamma_S/2)$,  
$|\Delta_L^{}| = |\Delta_R^{}|$ $(\equiv \Delta)$,
$\Gamma_S^{}= 2.0 \Delta$,   
  $\theta_R^{} = -\theta_L^{}$ $(\equiv \phi/2)$, 
and $\Gamma_N=0$. In this case 
the critical current is given by $J_C= e\Delta/\hbar$. 
} 
\label{Fig:results_2SC_J_dd}
\end{figure}

Figure \ref{Fig:results_2SC_J_dd} shows  
the Josephson current $J$ and the SC correlation 
$\langle d_\uparrow^{} d_\downarrow^{} \rangle$ 
in the impurity site as functions of $\phi$ 
 for several values of $U$. 
These ground-state averages vary discontinuously at the QPT, 
and take small negative values 
 in the magnetic doublet ground state for finite SC gaps $\Delta$. 
These two  expectation values in the doublet state are  
determined by the quasiparticle excitations 
above the SC energy gap $|\omega|>\Delta$, 
and vanish in the limit of $\Delta \to \infty$. 
The small negative values in the doublet state can be 
explained, for instance,  using the perturbation expansion 
with respect to $1/\Delta$ from 
the large  gap limit.\cite{Florens}

\section{Large SC gap limit}
\label{sec:large_sc_gap}

 We consider the large gap limit, $|\Delta_{L/R}| \to \infty$,  
in the following 
since important features of the interplays between the Kondo 
effect and superconductivity in the Y-junction can be observed    
in this case although the quasiparticle excitations 
to the continuum-energy region above the SC gap have been projected out.  
For instance, 
the Andreev resonance state emerging inside the SC gap remains near the 
Fermi level, and thus the essential physics of  
the low-energy transport can be extracted from this case.
Specifically, this limit describes reasonably the situation 
where the gap is much  greater than the other energy scales, namely 
 $|\Delta_{L/R}^{}| \gg 
\max(\Gamma_{L}, \Gamma_{R}, \Gamma_N,  U, |\epsilon_d|)$.

In  the limit of $|\Delta_{L/R}| \to \infty$,
the Hamiltonian $H$ can be mapped exactly 
onto a single-channel model,\cite{MartinYeyati,Affleck,TanakaNQDS}
\begin{align}
\mathcal{H}_\mathrm{eff}^{} =& \ \mathcal{H}_{dS} + H_{d} + H_N + H_{T,N}\;, 
\label{eq:Hami_seri_eff}
\\
\mathcal{H}_{dS}^{} = & \   
\Delta_{d}^{} \,
d^{\dag}_{\uparrow }d^{\dag}_{\downarrow }
+
\Delta_{d}^* \, 
d^{}_{\downarrow }
d^{}_{\uparrow }
\;, 
\label{Hamieff_S} \\
\Delta_d \equiv & 
\  \Gamma_R e^{i\theta_{R}^{}} + \Gamma_L e^{i\theta_{L}^{}} 
\, = \, |\Delta_d|\,e^{i\theta_{d}^{}}\;.
\label{eq:Delta_d}
\end{align}
Thus, the SC proximity effect becomes static in this case, and 
can be described by an additional term $\mathcal{H}_{dS}$ 
with the pair potential $\Delta_{d}$ penetrating into the QD.
This term emerges as the Cooper pairs can be transferred between    
the dot and the SC leads even for large SC gaps 
whereas  the unpaired quasiparticles cannot. 
The amplitude of $\Delta_d$ depends on the Josephson phase, 
and decreases as $\phi$ increases,  
\begin{align}
|\Delta_d| = & \  
\Gamma_S\,\sqrt{1 -\mathcal{T}_0 \sin^2\left(\phi/2\right)}
\;, 
\label{eq:Delta_d_def}
\\
\Gamma_S \equiv&\  \Gamma_R + \Gamma_L \;, 
\qquad \ \ 
\mathcal{T}_0 \,\equiv \, 
\frac{4\Gamma_R^{\phantom{0}}\Gamma_L^{\phantom{0}}}
{(\Gamma_R^{\phantom{0}}+\Gamma_L^{\phantom{0}})^2} 
\;.
\end{align}
Specifically for the symmetric coupling $\Gamma_R=\Gamma_L$,
the transmission probability in the normal-state case 
 takes the value  $\mathcal{T}_0=1$, 
and the  amplitude is  given simply by 
$|\Delta_d| = \Gamma_S\,\cos\left(\phi/2\right)$ for $-\pi<\phi\leq \pi$.
%


\subsection{Bogoliubov particles}
\label{subsec:bogo}

 The effective Hamiltonian $\mathcal{H}_\mathrm{eff}^{}$ 
can be transformed into an asymmetric  
Anderson model for the Bogoliubov particles, 
the total number of which 
is conserved.\cite{TanakaNQDS,TanakaNTDQDS,sc2}
In order to carry this out, 
we rewrite  $\mathcal{H}_\mathrm{eff}^{}$ such that 
\begin{align}
 \mathcal{H}_\mathrm{eff}^{} 
\,=&\, 
\left[
  d_{\uparrow}^{\dagger} ,
  d_{\downarrow}^{} 
 \right] 
\left[ \,
 \begin{matrix}
 \xi_d &  \Delta_d \cr
 \Delta_d^* & -\xi_d
 \end{matrix}
 \, \right]  
 \left[
 \begin{matrix}
  d_{\uparrow}^{\phantom{\dagger}} \cr
  d_{\downarrow}^{\dagger} \rule{0cm}{0.5cm}\cr
 \end{matrix}
 \right] 
+ \frac{U}{2} \left(n_{d}-1\right)^2  
\nonumber \\
&\  + 
\sum_{j=-1}^{\infty} \sum_{\sigma} 
t_{j}^{\phantom{0}} \! 
\left(
f^{\dagger}_{j+1\sigma}
f^{\phantom{\dagger}}_{j\sigma}
+
f^{\dagger}_{j\sigma} 
f^{\phantom{\dagger}}_{j+1\sigma}
\right) 
. 
\label{eq:Hami_seri_eff2}
\end{align}
Here, the summation over $j$ describes the $H_{T,N}+H_{N}$ part 
with  $f_{-1\sigma} \equiv d_{\sigma}$,
$f^{}_{0 \sigma} \equiv \psi_{N,\sigma}^{}$,
and $t_{-1} \equiv v_N^{}$.  
The operators $f_{j\sigma}$ for $j \geq 0$ correspond to 
a Wannier basis set for the conduction band. 
The explicit expression for 
$f^{}_{j \sigma}$ and $t^{}_{j}$ can be generated successively  
from the initial operator $f^{}_{0 \sigma}$ 
and the energy spectrum $\epsilon_k$ of the conduction band,
carrying out the Householder transformation.\cite{Hewson1}
Therefore, no approximation has been made 
to obtain Eq.\ \eqref{eq:Hami_seri_eff2}  
from Eq.\ \eqref{eq:Hami_seri_eff}.

The effective Hamiltonian $\mathcal{H}_\mathrm{eff}^{}$ has 
a global U(1) symmetry 
in the Nambu pseudo-spin space along 
the direction $\bm{n} 
\propto (|\Delta_d|\cos \theta_d,\,-|\Delta_d|\sin \theta_d,\,\xi_d)$.
Thus, one can choose this direction $\bm{n}$ to be 
the quantization axis, carrying out a pseudo-spinor rotation,   
\begin{align}
&\left[
 \begin{array}{c}
  \gamma_{j\uparrow}^{\phantom{\dagger}} \\
  (-1)^{j+1} \gamma_{j\downarrow}^{\dagger}
 \end{array}
\right]
=
\bm{U}^\dagger 
\left[
 \begin{array}{c}
  f_{j\uparrow}^{\phantom{\dagger}} \\
  (-1)^{j+1} f_{j\downarrow}^{\dagger}
 \end{array}
\right]\;,
 \label{eq:Bogo_A}
\end{align}
where
%
%
\begin{align}
%
& \bm{U} =
\left[ \,
 \begin{matrix}
   \  e^{i\frac{\theta_{d}^{}}{2}} & \,0 \cr
 0 &    e^{-i\frac{\theta_{d}^{}}{2}} 
 \end{matrix}
 \, \right]  
\left[ \,
 \begin{matrix}
   \  
 \cos \frac{\Theta_B^{}}{2} & 
\, -
\sin \frac{\Theta_B^{}}{2} \cr
 \sin \frac{\Theta_B^{}}{2} & 
\rule{0cm}{0.5cm} 
 \quad   
\cos \frac{\Theta_B^{}}{2}
 \end{matrix}
 \, \right] \;,  
\label{eq:Bogo_B}
\\ 
&
\cos \frac{\Theta_B}{2} =  
\sqrt{\frac{1}{2}\left(1+\frac{\xi_{d}}{E_A}\right)} \;, 
\qquad 
\cos \Theta_B =  \frac{\xi_d}{E_A} \;, 
\\
&
\sin \frac{\Theta_B}{2} =  
\sqrt{\frac{1}{2}\left(1-\frac{\xi_{d}}{E_A}\right)} \;,
\qquad
\sin \Theta_B =  \frac{|\Delta_{d}|}{E_A} \;,
\\
&E_A \equiv \sqrt{\xi_{d}^2+|\Delta_{d}|^{2}}
\;.
\label{eq:E_d}
\end{align}
%
%
Then, 
in terms of the Bogoliubov particles $\gamma_{j\sigma}^{\phantom{\dagger}}$, 
the effective Hamiltonian takes the form
\begin{align}
&\!\!\!
\mathcal{H}_\mathrm{eff}^{}=
E_A 
\left(n_{\gamma, -1}^{} -1
\right)+ \frac{U}{2}\left(n_{\gamma,-1}-1\right)^2
\nonumber\\
&
\qquad 
+\sum_{j=-1}^{\infty} \sum_{\sigma}
t_{j}^{\phantom{0}} 
\left(
\gamma_{j+1\sigma}^{\dagger}\gamma_{j\sigma}^{\phantom{\dagger}}
+\textrm{H.c.}
\right) .
\label{eq:after-Bogo}
\end{align}
Here,  $n_{\gamma,j}^{} \equiv \sum_{\sigma} 
 \gamma_{j\sigma}^{\dagger}\gamma_{j\sigma}^{\phantom{\dagger}}$,
and $\epsilon_d^\mathrm{eff} \equiv E_A- U/2$ corresponds to 
a bare impurity level for the Bogoliubov particles. 
This representation of the Hamiltonian 
clearly shows that the total number of Bogoliubov particles, 
\begin{align}
\mathcal{N}_\gamma \equiv \sum_{j=-1}^{\infty} 
n_{\gamma,j}^{} 
 \;,
\label{eq:Friedel}
\end{align}
is conserved in the large gap limit 
as a result of the global U(1) symmetry. 
Furthermore, the Friedel sum rule holds 
\begin{align}
\langle n_{\gamma, -1} \rangle 
\,  = \  \frac{2}{\pi}\,\delta \;, 
\label{MBogo} 
\end{align}
where $\delta$ is the phase shift of  
the Bogoliubov particles.\cite{Langer,Shiba}
Note that even in the case of $\xi_d=0$ where 
$\mathcal{H}_\mathrm{eff}^{}$ in the form of 
Eq.\ \eqref{eq:Hami_seri_eff2} has an electron-hole symmetry, 
 the Bogoliubov particles do {\it not\/} 
have the particle-hole symmetry 
in the sense that $E_A \neq 0$, 
and thus $\langle n_{\gamma, -1} \rangle\neq 1$, 
as long as $|\Delta_d|$ is finite.

At low energies  the interacting Bogoliubov particles, 
described by Eq.\ \eqref{eq:after-Bogo}, show 
Fermi-liquid behavior that is characterized by the renormalized parameters:  
\begin{align}
\delta \,\equiv& \  \cot^{-1} 
\left( \frac{\widetilde{E}_A}{\widetilde{\Gamma}_N} \right ) 
\;, \quad \ 
z^{-1} \, = \,  1 -
\left.\frac{\partial \Sigma(\omega)}{\partial \omega}
\right|_{\omega=0} , 
\label{eq:z_factor}
\\
 \widetilde{\Gamma}_{N} \equiv & \ 
z \,\Gamma_{N} \;, 
\qquad \qquad \quad 
\widetilde{E}_A \equiv  \,  
z \left [\,
E_A + \Sigma(0) \,\right ] ,  
\\
\widetilde{U} 
\,\equiv &\ z^2 
\Gamma_{\uparrow\downarrow;\downarrow\uparrow}^{}(0,0;0,0) .
\label{eq:U_ren}
\end{align}
Specifically, the Kondo energy scale can be deduced from 
the renormalization factor as $T_K=\pi \widetilde{\Gamma}_{N}/4$.
Furthermore, from the residual interaction $\widetilde{U}$ 
 between the quasiparticles,   
the Wilson ratio $R\,$  can be deduced through 
\begin{align}
R \,\equiv\, 1 + 
\frac{\widetilde{U}}{\pi\widetilde{\Gamma}_N}
\,\sin^2 \! \delta \;.
%
\end{align}
We calculate these renormalized parameters with the NRG 
through the convergence of the finite-size energy spectrum 
near the fixed point.\cite{KWW2,Hewson2}
In Eq.\ \eqref{eq:z_factor}-\eqref{eq:U_ren},
 $\Sigma(\omega)$ and 
$\Gamma_{\uparrow\downarrow;\downarrow\uparrow}^{}
(\omega_1,\omega_2;\omega_3,\omega_4)$ 
are  the self-energy and  vertex function, respectively,
for the Bogoliubov particles, the retarded Green's function 
for which is defined by
\begin{align}
G_{\gamma}^{}(\omega) 
\, =\,  -i\int_0^\infty \! dt \, e^{i\omega t} 
\left\langle 
\left\{
\gamma_{-1,\sigma}^{}(t)\,,\, \gamma_{-1,\sigma}^{\dagger}
\right\} \right\rangle \;. 
\end{align}
Here,  the spin suffix  $\sigma$ is suppressed 
on the left-hand side because the 
Green's function for $\sigma=\uparrow$ and that 
for $\downarrow$ are identical due to the SU(2) symmetry 
for the real spin.
The retarded Green's function for the electrons 
on the dot can be deduced from 
$G_{\gamma}^{}(\omega)$ 
via the inverse Bogoliubov transform, 
\begin{align}
G_{dd}^{}(\omega) 
 \equiv & \  -i\int_0^\infty \! dt \, e^{i\omega t} 
\left\langle 
\left\{
d_{\sigma}^{}(t)\,,\, d_{\sigma}^{\dagger}
\right\} \right\rangle 
\nonumber 
\\
\Rightarrow& \ 
G_{\gamma}^{}(\omega) 
\,\cos^2 \frac{\Theta_B}{2} 
- \left\{G_{\gamma}^{}(-\omega) \right\}^* 
\sin^2 \frac{\Theta_B}{2} .
\label{eq:Gdd_Ggam}
\end{align}

\subsection{Conductance and Current}

The low-energy transport, 
deduced from  Eq.\ \eqref{eq:after-Bogo},  
can also be described by the local Fermi liquid theory.
At $T=0$,  the occupation number of the electrons 
$\langle n_{d} \rangle$ and the SC pair correlation 
$\langle d_{\downarrow}^{} d_{\uparrow}^{} \rangle$ 
in the QD are determined by the occupation of 
the Bogoliubov particles $\langle n_{\gamma,-1} \rangle
=2\delta/\pi$ defined in Eq.\ \eqref{eq:Friedel},  
and the Bogoliubov angle $\Theta_B$, 
\begin{align}
\langle n_{d} \rangle -1
=& \ 
 \Bigl(\langle n_{\gamma, -1} \rangle -1\Bigr)        
 \cos \Theta_B^{} 
\;,
\label{eq:M_sDQD}
\\
\langle d_{\downarrow}^{} d_{\uparrow}^{} \rangle
\ =&  \ 
\frac{1}{2}
\Bigl(\langle n_{\gamma, -1} \rangle -1\Bigr)
 \,e^{i \theta_{d}^{}} \sin \Theta_B^{} \;. 
\label{eq:K_sDQD}
\end{align}
Note that $\cos \Theta _B^{} =\xi_{d}/E_{A}$, 
  $e^{i \theta_{d}^{}} \sin \Theta_B^{}=\Delta_{d}/E_{A}$, and   
 the phase of $\langle d_{\downarrow}^{} d_{\uparrow}^{} \rangle$ 
is given by $\theta_d$, which 
coincides with the phase of the local gap $\Delta_d$.
The occupation number of the electron  
$\langle n_{d} \rangle -1$ and 
 $\langle d_{\downarrow}^{} d_{\uparrow}^{} \rangle$ 
correspond to 
the $z$ component and the projection on the $x-y$ plane  
of the local pseudo-spin moment induced on the impurity site.\cite{sc2}  
Specifically, in the electron-hole symmetric case 
$\xi_d =0$, the local level is given by $E_A = |\Delta_d|$ and 
thus the Bogoliubov angle is locked at $\Theta_B^{}=\pi/2$.

The dc conductance  $g_{NS}^{}$  due to the Andreev scattering 
between the dot and the normal lead 
can also be expressed, at $T=0$, in terms of 
the phase shift $\delta$ and Bogoliubov angle $\Theta_B$,\cite{TanakaNQDS}  
\begin{align}
g_{NS}^{}
 = &\
4\,
\Gamma_N^2\,
\Bigl| 
\left\{\bm{G}_{dd}^{r}(\omega=0) \right\}_{12}^{} 
\Bigr|^2  \nonumber \\
 =  & \ 
\frac{4e^2}{h}
 \, \sin^2 \Theta_B^{}  \, \sin^2 2 \delta  .
\label{eq:Cond}
\end{align}
Here, $\{\bm{G}_{dd}^{r}(\omega)\}_{12}=
\langle \! \langle d_{\uparrow}^{} ; d_{\downarrow}^{}
\rangle \! \rangle_{\omega}^{}$ 
is the off-diagonal (anomalous) component of the retarded Green's function 
in the Nambu pseudo-spin formalism, 
the corresponding Matsubara function of which is defined  
in Appendix  \ref{sec:appendix_Green}.
Therefore, the zero-temperature conductance $g_{NS}^{}$ 
is determined by the value at the Fermi level $\omega=0$.
The argument $2\delta$ in Eq.\ \eqref{eq:Cond}
appears as a difference between the phase shift for 
the quasi-particle $+\delta$ and that for the quasi-hole $-\delta$.

In the large gap limit, 
the Josephson current flowing through the dot  
can also be expressed in terms of the 
Bogoliubov angle $\Theta_B^{}$ and the phase shift $\delta$, 
or $\langle n_{\gamma, -1} \rangle$ 
[see Appendix \ref{sec:appendix_Green}],
\begin{align}
\langle J \rangle 
\, = & \, 
J_C^{} 
\,
 \mathcal{T}_0 \ 
\frac{\,  
 \bigl|\langle n_{\gamma, -1} \rangle -1\bigr|        
\, \sin \Theta_B^{}\,\sin \phi \,
}
{2\sqrt{1 - \mathcal{T}_0\sin^2\left(\phi/2\right)}}
\,
\;.
\label{eq:JosephsonCurrentY-junction}
\end{align}
 Here, $J_C \equiv {e \Gamma_S}/{\hbar}$ is the critical current. 
Note that $\langle J \rangle $ shows 
a non-sinusoidal dependence on $\phi$ in general 
because  $\Theta_B^{}$, $\langle n_{\gamma, -1} \rangle$, and 
the denominator of Eq.\ \eqref{eq:JosephsonCurrentY-junction} 
that arises through $|\Delta_d|$ vary as functions of $\phi$.


\begin{figure}[t] 
 \leavevmode
\begin{minipage}{1\linewidth}
\includegraphics[width=0.9\linewidth]{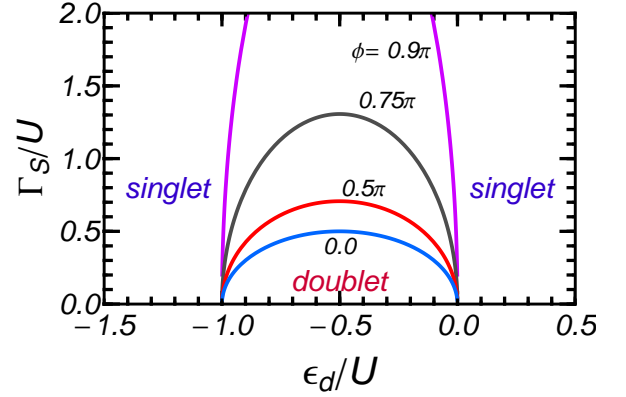}
\end{minipage}
\caption{
(Color online) 
Ground-state phase diagram in 
the limit of $\Gamma_N \to 0$ and $|\Delta_{L/R}| \to \infty$ 
is plotted in  $\epsilon_d$ vs $\Gamma_S^{}$ plane 
for several values of $\phi$.
The couplings between the QD and  SC leads is chosen to be symmetric 
 $\Gamma_L^{} = \Gamma_R^{}$ $(\equiv \Gamma_S^{}/2)$. In this case 
the local SC gap 
is given by  $\Delta_d=\Gamma_S^{} \cos \phi/2$ for $|\phi|\leq \pi$.
The ground state is a singlet due to the local Cooper pairing 
outside the semi ellipse of    
$\sqrt{(\epsilon_d +U/2)^2 +|\Delta_d|^2} =U/2$.   
Inside this semi ellipse the ground state is 
a magnetic doublet, which for finite $\Gamma_N$ changes to a Kondo singlet 
due to the conduction-electron screening of the local moment.
 Note that at  $\epsilon_d = -0.5U$ 
the system has the electron-hole symmetry, 
and the results for this case is given 
in Fig.\ \ref{Fig:results_2SC_GS} at finite SC gaps.  
} 
\label{fig:phase_diagram_atomic_limit} 
\end{figure}

\subsection{Kondo singlet vs Local Cooper pairing}

The ground state of the asymmetric Anderson model, given in 
 Eq.\ \eqref{eq:after-Bogo},  can be 
classified according to the fixed points of the NRG.\cite{KWW2,Hewson1}
Among them the {\it strong-coupling\/} fixed point 
describes the Kondo singlet, 
for which the impurity site is singly occupied.  
The {\it frozen-impurity\/} fixed point describes a 
different situation, where the impurity level is 
far away from the Fermi level and 
the impurity site becomes empty or doubly occupied.
In our case, the {\it frozen-impurity\/} fixed point 
is defined with respect to Eq.\ \eqref{eq:after-Bogo} 
for the Bogoliubov particles, 
and thus this describes a singlet state 
by a local Cooper pairing that  
consists of a linear combination of 
the empty and doubly occupied impurity states 
of the electrons represented in Eq.\ \eqref{eq:Hami_seri_eff2}.
We refer to this fixed point as 
 {\it local Cooper pairing\/} in the following.

The ground state of 
$\mathcal{H}_\mathrm{eff}^{}$ 
varies continuously between 
 the Kondo singlet and the local-Cooper pairing,  
depending on the Hamiltonian parameters $E_A$, $U$, and $\Gamma_N$.
A rough sketch of the ground-state phase diagram can be 
obtained quickly from that 
in the atomic limit $\Gamma_{N}^{\phantom{0}} \to 0$, 
where the normal lead is disconnected.
In this limit the dot is occupied by 
a single Bogoliubov particle with spin $1/2$ for $E_{A}<U/2$ whereas 
the dot is empty with no Bogoliubov particle for $E_{A}>U/2$.
Thus, the phase boundary is given by a simple semi ellipse of  
$\sqrt{\xi_{d}^2+|\Delta_{d}|^{2}}=U/2$,
which is plotted in Fig.\ \ref{fig:phase_diagram_atomic_limit} 
for several values of $\phi$. 
The ground state is a singlet due to the local Cooper pairing 
outside the semi ellipse whereas  
inside the semi-elliptic boundary  
 the ground state is a doublet and thus  
the local moment arises in this limit of $\Gamma_N \to 0$.  
However, the local moment is screened 
when the normal lead is connected $\Gamma_N \neq 0$, 
and then the ground state inside the semi ellipse 
changes to the Kondo singlet. 
The coupling to the normal lead also changes the sharp transition 
at the boundary to a continuous crossover between  
the local-Cooper-pairing singlet and the Kondo singlet.\cite{TanakaNQDS}

There are further quantitative corrections  when the SC gap $\Delta$ is finite.
This was also examined for the $\Gamma_N=0$ case, 
using the NRG.\cite{YoshiokaOhashi,sc3}
The results showed that the region of the magnetic doublet state 
becomes small as $\Delta$ decreases.
This is because also the excited states in the SC leads 
 above the gap $|\omega| > \Delta$ 
  contribute to the screening of the local moment for finite $\Delta$.
%


\section{NRG Results} \label{sec:result}

In this section, we provide the NRG results for 
the ground-state properties of the Y-junction 
in the large gap limit $|\Delta_{L/R}| \to \infty$.

\subsection{Spectral function}

We first of all discuss the impurity spectral function 
for electrons  
 $A_{dd} =  (-1/\pi)\, 
\mathrm{Im}\, G_{dd}^{}$ that  
can be deduced from the one 
for the Bogoliubov particles
$A_{\gamma} 
=  (-1/\pi) \,\mathrm{Im}\, G_{\gamma}^{}$, 
\begin{align}
A_{dd}^{}(\omega) \, 
= & \ 
 A_{\gamma}(\omega) 
\,\cos^2 \frac{\Theta_B}{2} 
+ A_{\gamma}(-\omega) 
\,\sin^2 \frac{\Theta_B}{2} \;.
\label{eq:A_dd_gamma}
\end{align}
Specifically, the spectrum is  
symmetric $A_{dd}^{}(-\omega)=A_{dd}^{}(\omega)$  
 in the electron-hole symmetric case, where $\Theta_B =\pi/2$.   
Note that the single-electron spectrum  $A_{dd}^{}(\omega)$ 
consists of a superposition    
of a single-Bogoliubov-particle part $A_{\gamma}(\omega)$ and 
a single-Bogoliubov-hole part $A_{\gamma}(-\omega)$. 
This can be deduced from the Lehmann representation 
that is expressed in terms of 
the matrix element  $\langle m,N_b'| 
d_{\uparrow}^\dagger|\mathrm{GS},N_b \rangle$ 
between the ground state $|\mathrm{GS},N_b \rangle$ 
and an excited state $|m,N_b' \rangle$ 
of $\mathcal{H}_\mathrm{eff}$, where $N_b$ is 
an eigenvalue of the total number of the 
Bogoliubov particles $\mathcal{N}_\gamma$ defined in
 Eq.\ \eqref{eq:Friedel}.
This matrix element can be finite not only for  
the single-particle excitations with $N_b'=N_b+1$ 
but also single-hole excitations with $N_b'=N_b-1$ 
of the Bogoliubov particles because 
the electron $d_{\uparrow}^\dagger$ can be decomposed into 
 a superposition of the annihilation $\gamma_{-1,\uparrow}^\dagger$ 
and creation $\gamma_{-1,\downarrow}^{}$ of the Bogoliubov particles. 

The low-energy spectral weight is dominated by the renormalized 
Andreev resonances that appear in  $A_{dd}^{}(\omega)$ 
 as a pair of  quasiparticle peaks,  
\begin{align}
A_{dd}^{}(\omega) \, 
\simeq  \, \frac{z}{\pi} \left[
\,\frac{\widetilde{\Gamma}_N \cos^2 \frac{\Theta_B}{2}}
{\left( \omega - \widetilde{E}_A \right)^2_{} + \widetilde{\Gamma}_N^2} 
+ 
\,\frac{\widetilde{\Gamma}_N \sin^2 \frac{\Theta_B}{2}}
{\left(\omega + \widetilde{E}_A\right)^2_{} + \widetilde{\Gamma}_N^2} 
\right]
.
\label{eq:A_dd_quasiparticle}
\end{align}
The peak position $\pm \widetilde{E}_A$, width $\widetilde{\Gamma}_N$, 
and renormalization factor $z$ vary as the Coulomb repulsion increases 
from  $U= 0$, 
for which we have  $\widetilde{E}_A = E_A$, 
 $\widetilde{\Gamma}_N = \Gamma_N$, and $z = 1$.

\begin{figure}[t] 
 \leavevmode
\begin{minipage}{1\linewidth}
\includegraphics[width=0.75\linewidth]{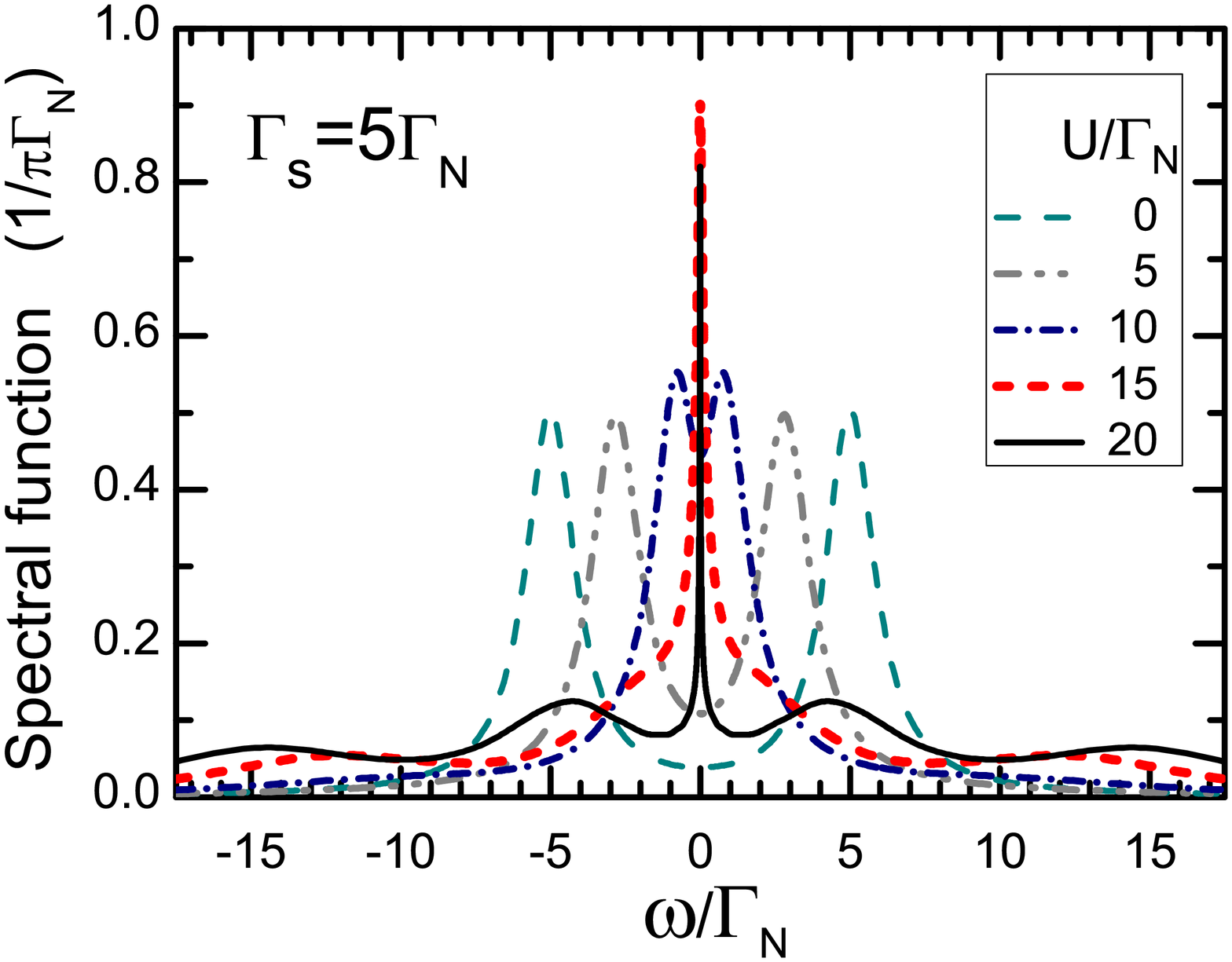}
\includegraphics[width=0.75\linewidth]{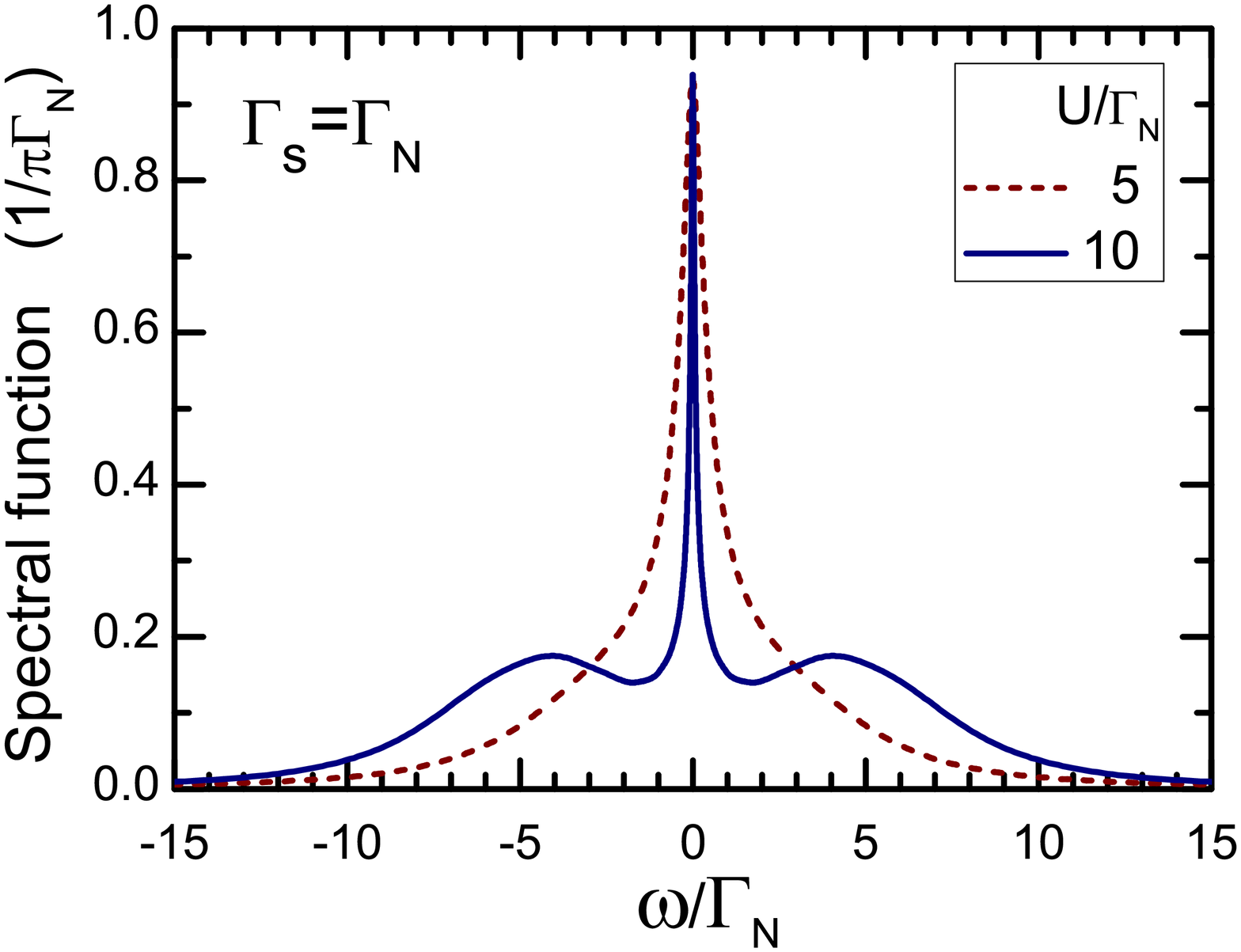}
\end{minipage}
\caption{
(Color online)
NRG results for the Spectral function 
$A_{dd}^{}(\omega)= (-1/\pi)\, \mathrm{Im}\, G_{dd,\sigma}^{}(\omega)$,
in the large gap limit $|\Delta_{L/R}| \to \infty$ at $\phi =0$, 
is plotted for several value of $U$ 
choosing the couplings to the SC leads such that 
(upper panel) $\Gamma_S/\Gamma_N=5.0$, 
and (lower panel) $\Gamma_S/\Gamma_N=1.0$. 
The other parameters are taken to be 
$\Gamma_L=\Gamma_R$ ($\equiv \Gamma_S/2$) 
assuming the electron-hole symmetric $\epsilon_d = -U/2$. 
In the present case the local SC gap in the impurity site is given 
by $\Delta_d=\Gamma_S^{}$, and $E_A=\Gamma_S^{}$. 
} 
\label{fig:spec_phi=0}
\end{figure}

The high-energy part of the spectral weight  away from the Fermi level 
can be inferred from the one in the atomic limit $\Gamma_N^{}=0$. 
For weak repulsions $E_A>U/2$, the ground state is a singlet, 
and  $A_{dd}^{}(\omega)$ has two peaks at $\omega= \pm(E_A-U/2)$. 
This is because in this case $A_{\gamma}(\omega)$ has a single peak 
at $\omega=E_A-U/2$, which moves towards the Fermi level as $U$ increases. 
On the other hand, for strong repulsions $E_A<U/2$ in the $\Gamma_N^{} \to 0$ limit, 
the ground state is a magnetic doublet, 
and then  $A_{dd}^{}(\omega)$  has 
four peaks emerging at $\omega = \pm \epsilon_\mathrm{UP}^{}$ 
and $\pm \epsilon_\mathrm{LW}^{}$. These peaks   
are caused by the excitations to 
the upper $\epsilon_\mathrm{UP}^{} \equiv U/2+E_A$ and lower 
$\epsilon_\mathrm{LW}^{} \equiv -(U/2-E_A)$ atomic peaks defined with respect  
to the Bogoliubov particles, and each of the two final states consists  
of a linear combination of an empty 
and doubly occupied states of the original electrons. 
These peaks correspond to the original Andreev bound states  
typically found in situations of a QD coupled only to a superconductor, 
and they are delta-functions in the limit $\Gamma_N^{}=0$. 
For finite superconducting gap their position and the occurrence of the ground 
state transition change quantitatively.\cite{sc3} In this situation in 
the doublet phase it is possible that the higher excitations 
$\pm \epsilon_\mathrm{UP}^{}$ are not found within the gap anymore.

Figure \ref{fig:spec_phi=0} shows the  
spectral function for $\phi=0$ 
in the electron-hole symmetric case, 
where $E_A= \Gamma_S^{}$. In the upper panel $A_{dd}^{}(\omega)$ in the case 
of relatively small $\Gamma_N^{}$ with $E_A=5\Gamma_N^{}$ is plotted for several values of 
the Coulomb repulsion $U/\Gamma_N^{}=0, \,5,\,10,\,15,\,20$. 
In these examples, at $U= 10 \Gamma_N^{}$ 
the bare parameter takes the value $E_A = U/2$, and thus the dash-dot line 
represents the results obtained at  
the crossover region between the Kondo and local SC singlet states.
The pair of  renormalized Andreev resonances 
at $\omega \simeq \pm \widetilde{E}_A^{}$, 
which correspond to the Kondo peak for the Bogoliubov particles 
described in Eq.\ \eqref{eq:A_dd_quasiparticle},
shift closer to the Fermi level $\omega=0$ as the Coulomb repulsion 
$U$ increases from $0$ to $2E_A$. 
Then, for $U>2E_A$  five peaks emerge 
as seen the curves for $U=15\Gamma_N^{}$  
and $20\Gamma_N^{}$. 
Among them, the central peak near 
the Fermi level $\omega=0$ corresponds to the Kondo 
resonance for the Bogoliubov particles, 
which appears for  $A_{\gamma}^{}(\omega)$  at small 
positive frequency $\omega\simeq \widetilde{E}_A^{}$. 
However, for the electron spectral function $A_{dd}^{}(\omega)$, 
 this peak at $\omega\simeq \widetilde{E}_A^{}$
and the counterpart for holes  
at $\omega \simeq -\widetilde{E}_A^{}$ 
 overlap to form a single peak at the Fermi energy $\omega=0$.
The other four peaks represent 
the electron and hole components of the excitations  
to the the upper and lower atomic levels of the Bogoliubov particles.
For instance, in the curve for $U=20\Gamma_N^{}$, 
the broad peak 
at $\omega \simeq -5\Gamma_N^{}$ and 
the one at $\omega \simeq +15\Gamma_N^{}$ correspond to 
 upper and lower atomic peaks appearing  
in  $A_{\gamma}(\omega)$, respectively.

The lower panel of Fig.\ \ref{fig:spec_phi=0} shows 
the spectral function, obtained at $E_A =\Gamma_N^{}$ 
where the bare Andreev level and the hybridization energy 
due to the coupling to the normal lead coincide.
The Coulomb interaction is chosen 
to be  $U = 5\Gamma_N^{}$ and $10\Gamma_N^{}$, 
so that $E_A<U/2$ for these two cases. 
As the coupling to the normal lead $\Gamma_N^{}$ is 
relatively large in these cases 
the sub peaks of the Coulomb oscillation 
are smeared especially in the curve for $U = 5\Gamma_N^{}$.
Nevertheless, in the other curve for $U = 10\Gamma_N^{}$, 
the sharp central peak and 
two sub peaks of the atomic nature are distinguishable.
The central peak emerges as a results of the superposition of 
the two renormalized Andreev level at $\omega \simeq \pm \widetilde{E}_A$ 
close to the Fermi level
whereas each of the sub peaks at $\omega\simeq \pm 4.0 \Gamma_N$ emerge 
as a sum of the sub peaks in  $A_{\gamma}(\omega)$ 
and that in the counterpart for the holes $A_{\gamma}(-\omega)$.

The results which we have presented in Fig.\ \ref{fig:spec_phi=0} 
have been the ones obtained at $\phi=0$, where there is no phase difference 
between the two SC order parameters. 
As we have chosen the parameters such 
that  $\xi_d =0$ and $\Gamma_L=\Gamma_R$, 
the dependence of  $A_{dd}(\omega)$ on $\phi$ enters  
only through the bare Andreev 
level $E_A = \Gamma_S \cos \phi/2$ in this case.
Therefore, the results capture essential features 
 common to the case for finite $\phi$. 
In the electron-hole asymmetry case $\xi_d \neq 0$,  
however,  $A_{dd}(\omega)$ is no longer 
a symmetric function of $\omega$.
The asymmetry in the $\omega$ dependence enters  
through  $\Theta_B$ as it deviates from $\pi/2$ 
for $\xi_d \neq 0$.

\subsection{Transport properties \& Fermi-liquid behavior}

In this subsection we present the NRG results for the ground-state properties 
of the Y-junction for symmetric 
coupling $\Gamma_L=\Gamma_R$ $(=\Gamma_S^{}/2)$, where $\mathcal{T}_0=1$. 
The Josephson phase is chosen such that 
$\theta_R = -\theta_L $ $(=\phi/2)$, 
which makes the local SC gap $\Delta_d$ and 
 pair correlation  
$\langle d_{\downarrow}^{} d_{\uparrow}^{} \rangle$ 
real $\theta_d=0$ as shown in Appendix.
We consider the electron-hole symmetric case 
$\xi_d=0$  in Sec.\ \ref{sec:e-h_symmetric_case}, 
and then discuss the gate voltage dependence  
varying $\xi_d$ in Sec.\ \ref{sec:e-h_asymmetric_case}.

\begin{figure}[t]

\begin{minipage}{1\linewidth}
 \leavevmode
\includegraphics[width=0.48\linewidth]{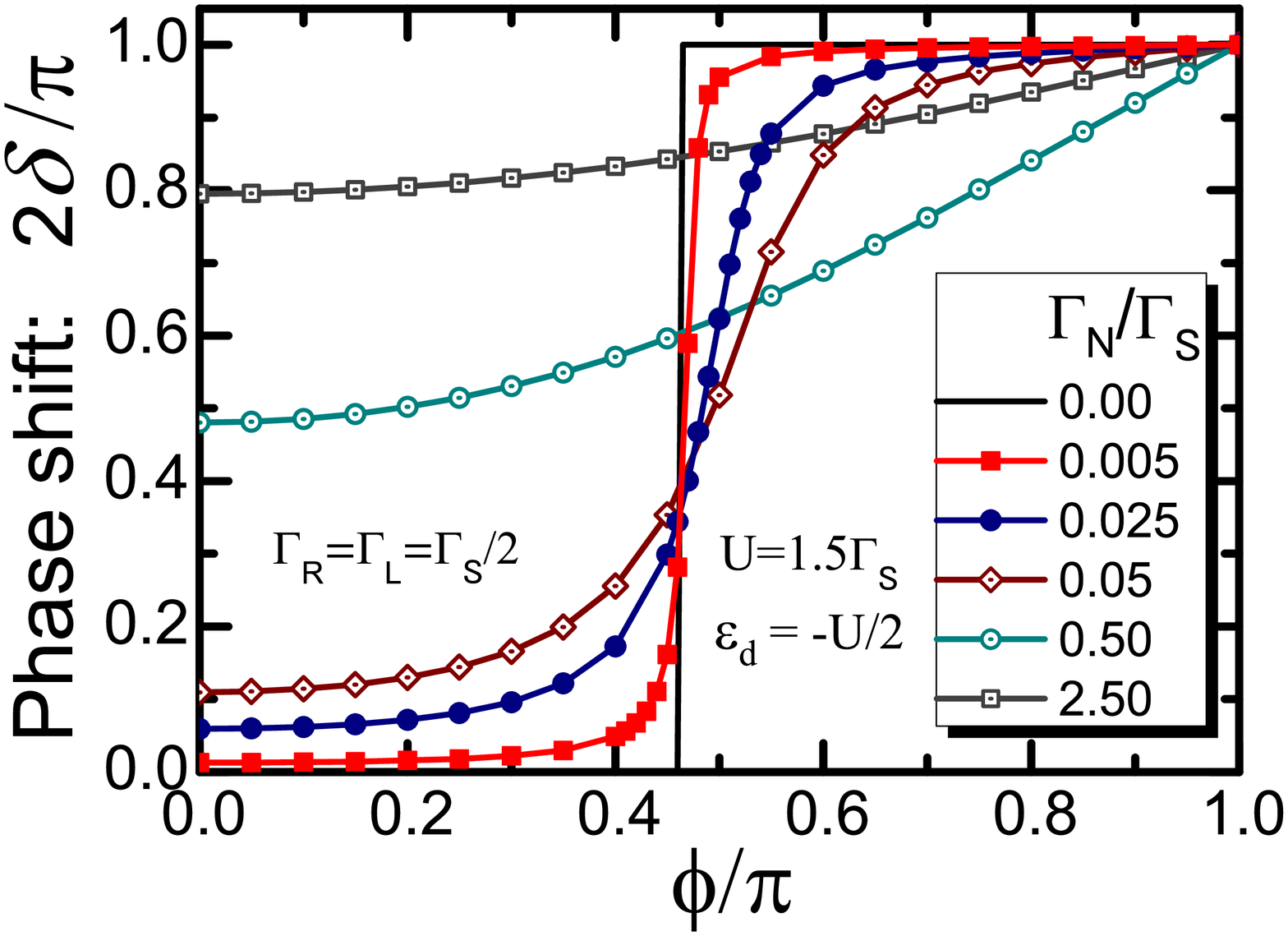}
\includegraphics[width=0.48\linewidth]{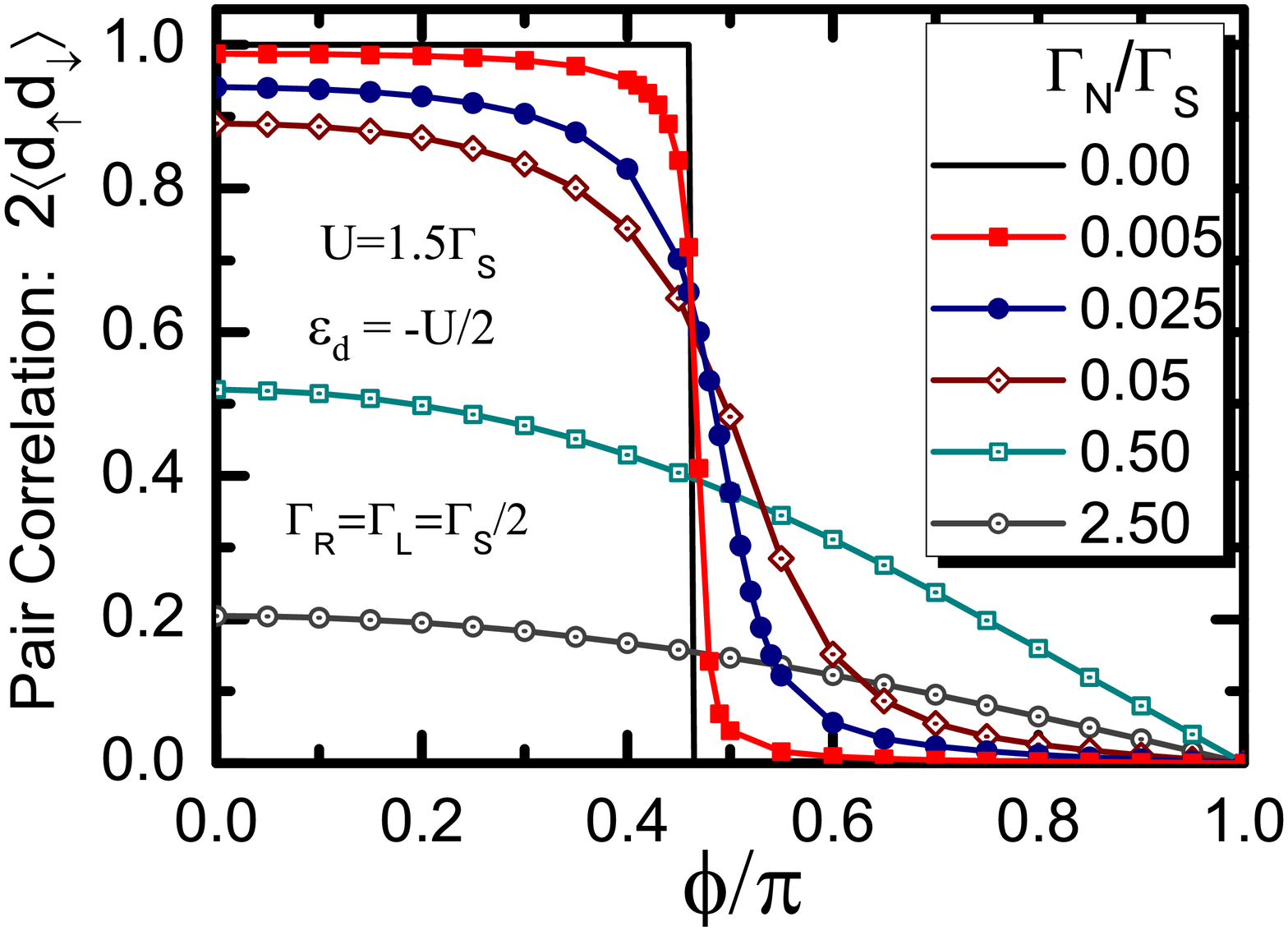}
\end{minipage}
\\
\vspace{0.3cm}
\begin{minipage}{1\linewidth}
\includegraphics[width=0.48\linewidth]{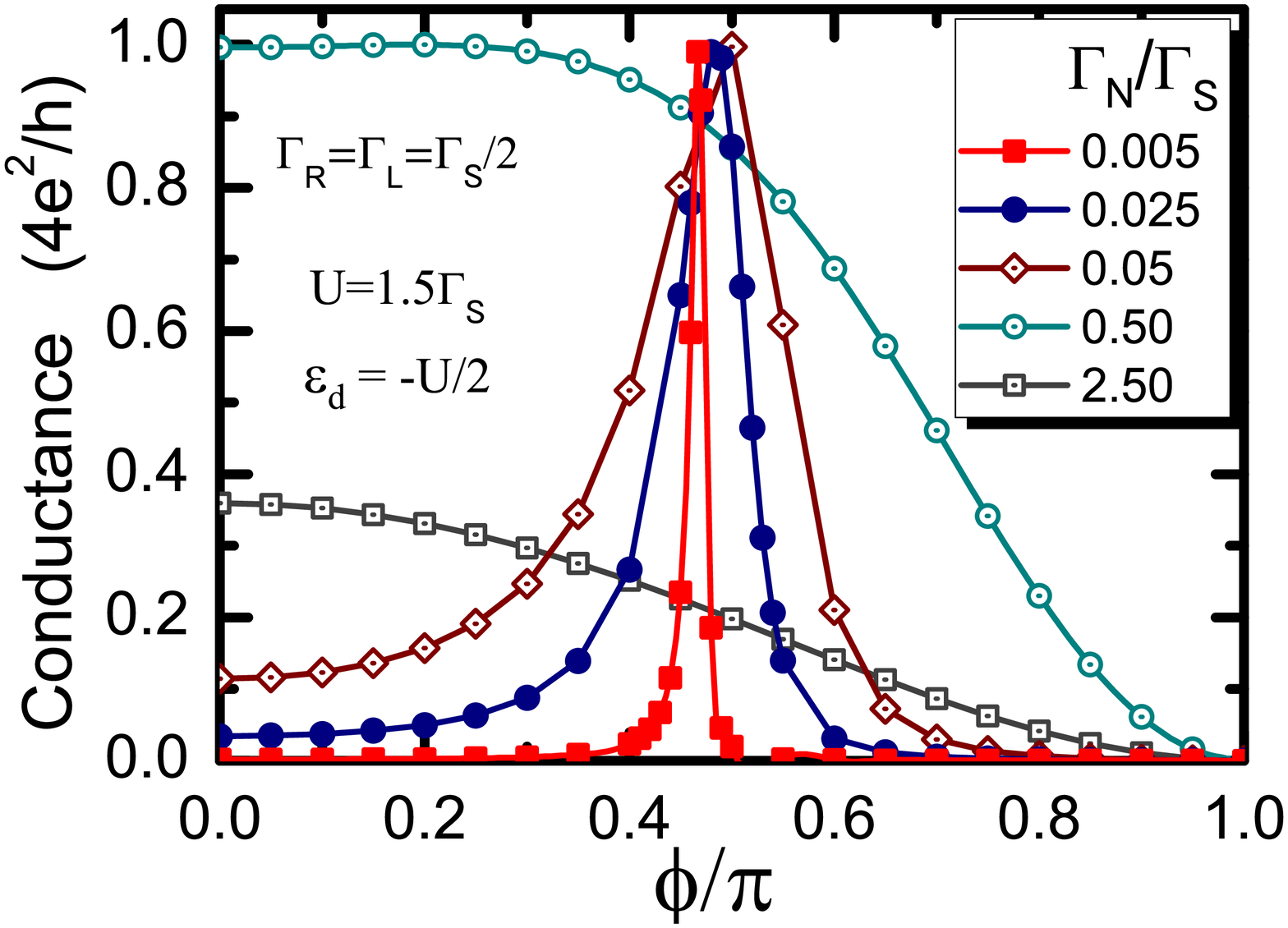}
\includegraphics[width=0.485\linewidth]{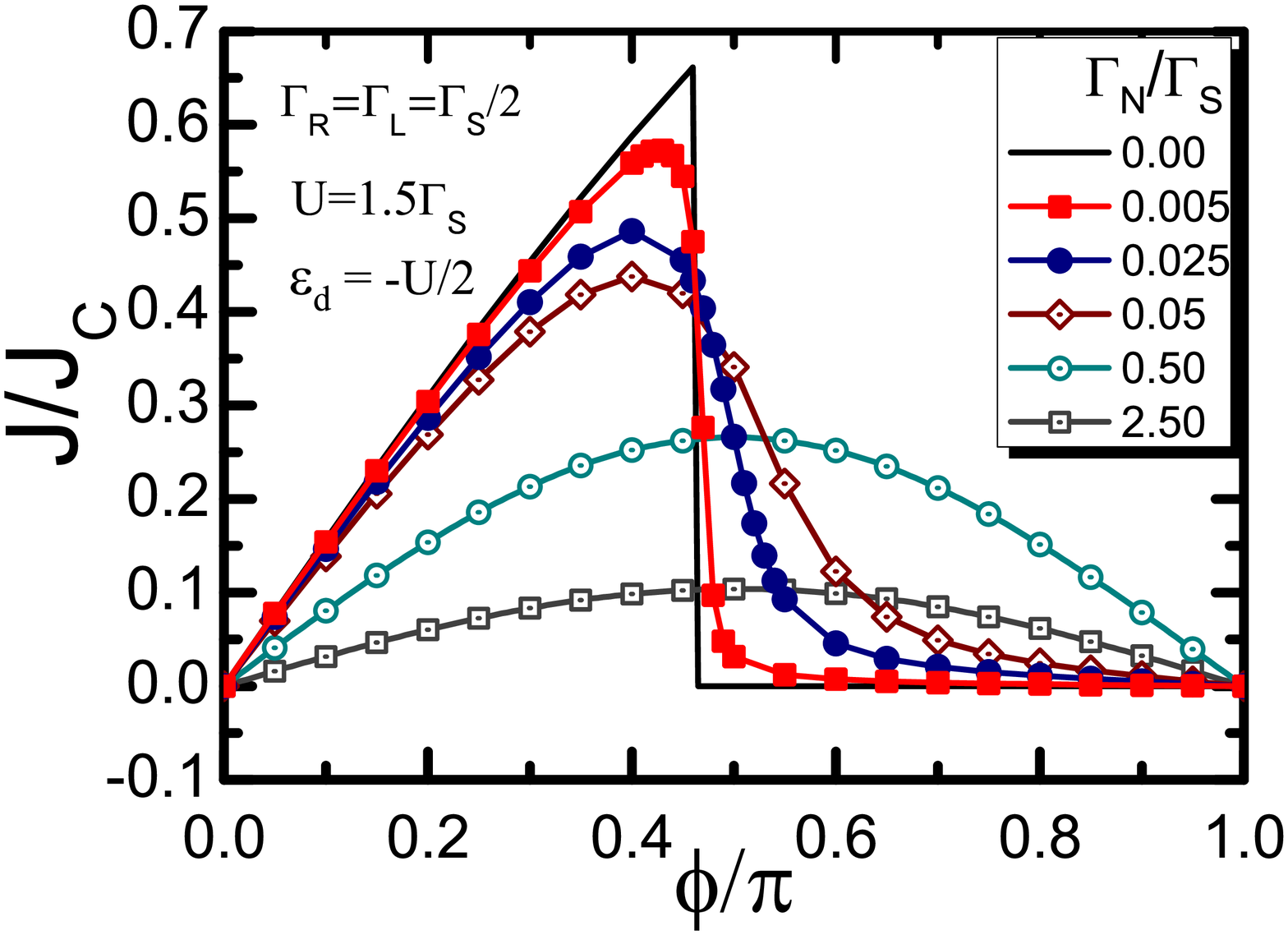}
\end{minipage}
\caption{
(Color online) 
Phase shift and some related ground-state averages    
are plotted vs Josephson phase $\phi$ 
 for several values of $\Gamma_N$ 
in the electron-hole symmetric case $\epsilon_d = -U/2$: 
(upper panel) phase shift $\delta$ 
and pair correlation  
$2\left|\langle d_{\downarrow}^{} d_{\uparrow}^{} \rangle\right|$,  
(lower panel)  conductance $g_{NS}^{}$ and  
Josephson current $J$ in units of $J_C = e\Gamma_S^{}/\hbar$.
The other parameters are chosen such that 
 $\Gamma_L=\Gamma_R$ ($= \Gamma_S/2$), 
and  $U=1.5\Gamma_S$ 
in the large gap limit $|\Delta_{L/R}| \to \infty$.  
}
\label{fig:YJ_cond_half}
\end{figure}

\begin{figure}[t]

\begin{minipage}{1\linewidth}
 \leavevmode
\includegraphics[width=0.75\linewidth]{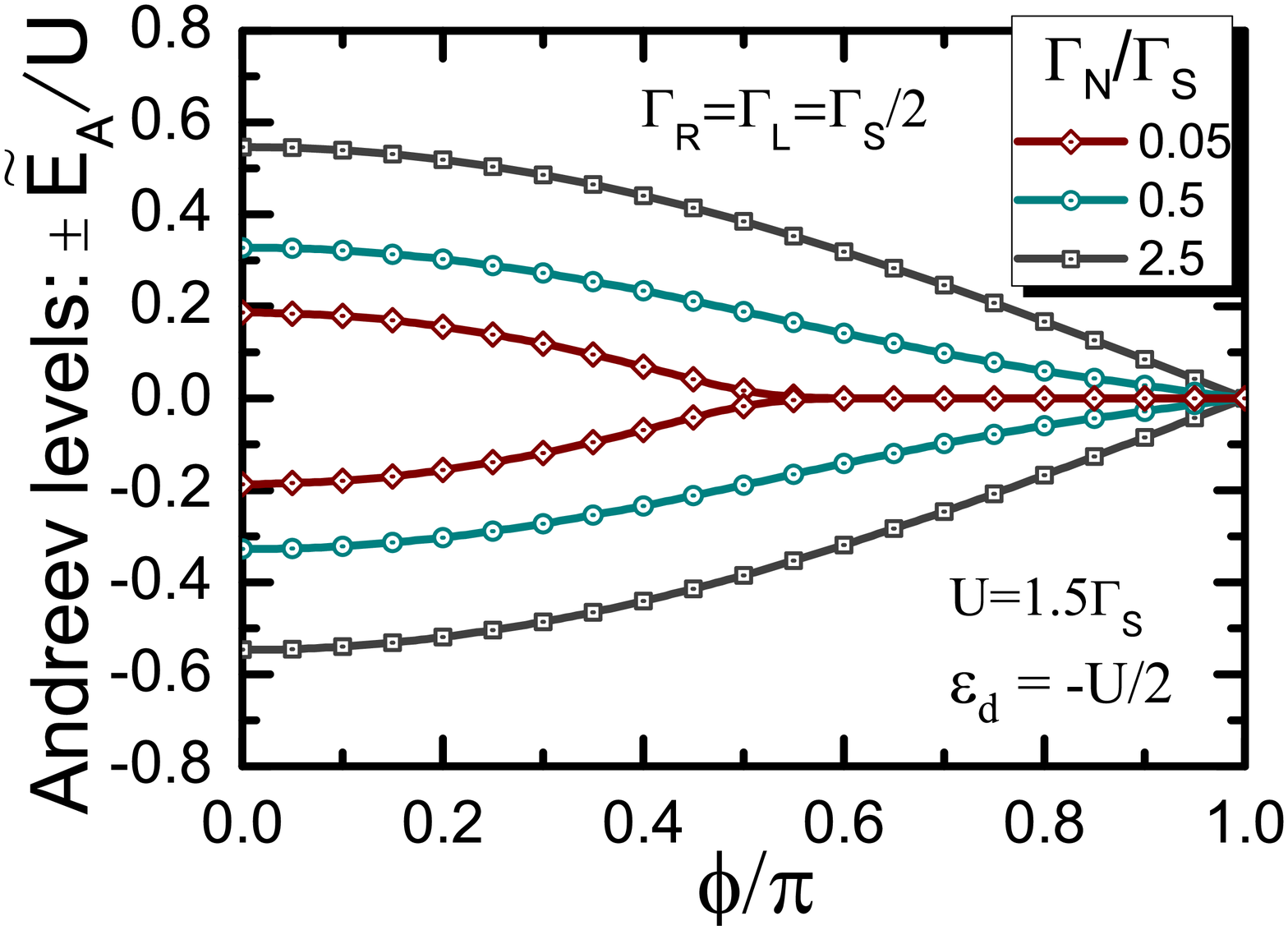}
\end{minipage}
\\ \vspace{0.3cm}
\begin{minipage}{1\linewidth}
 \leavevmode
\includegraphics[width=0.477\linewidth]{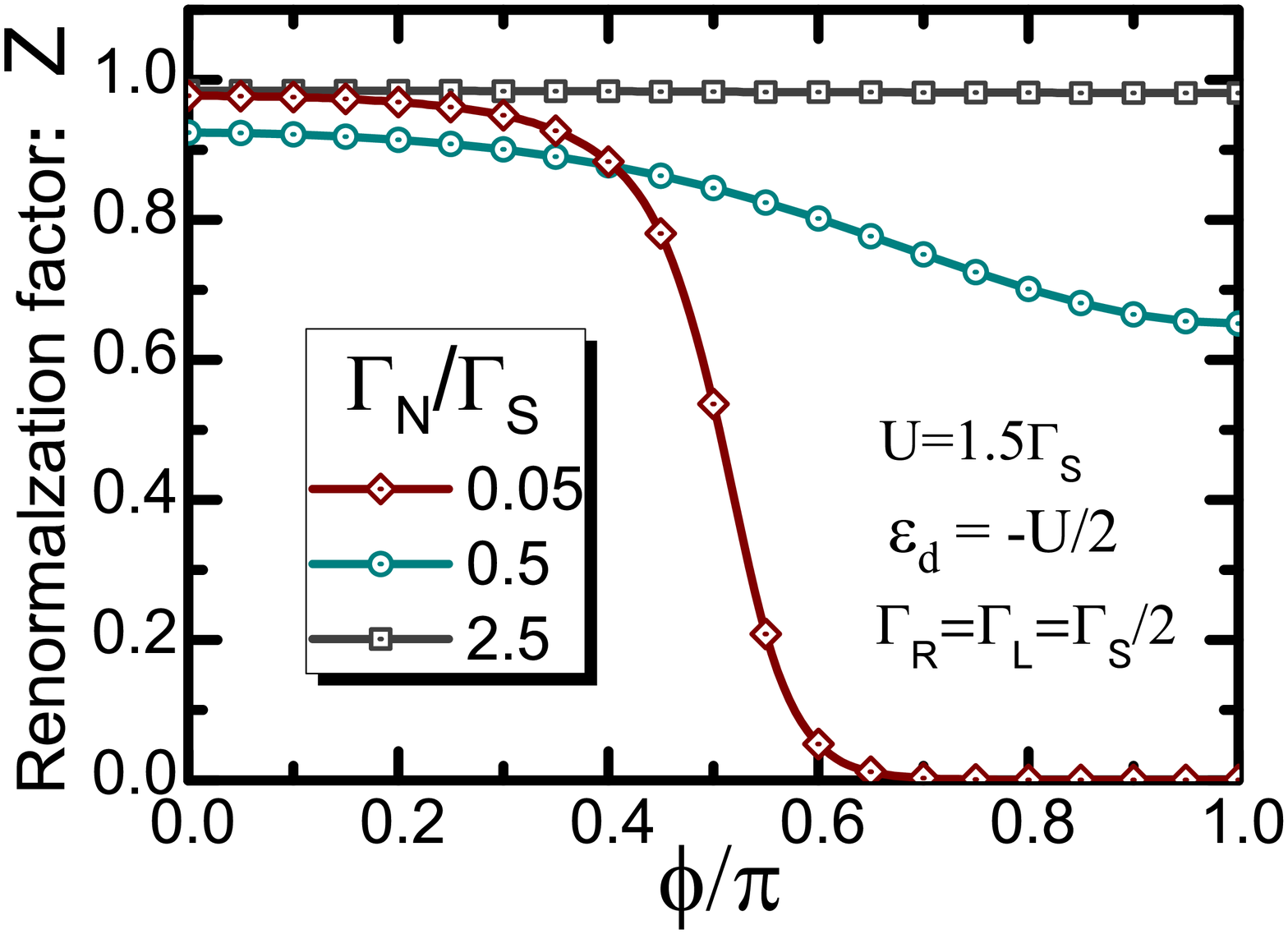}
\includegraphics[width=0.494\linewidth]{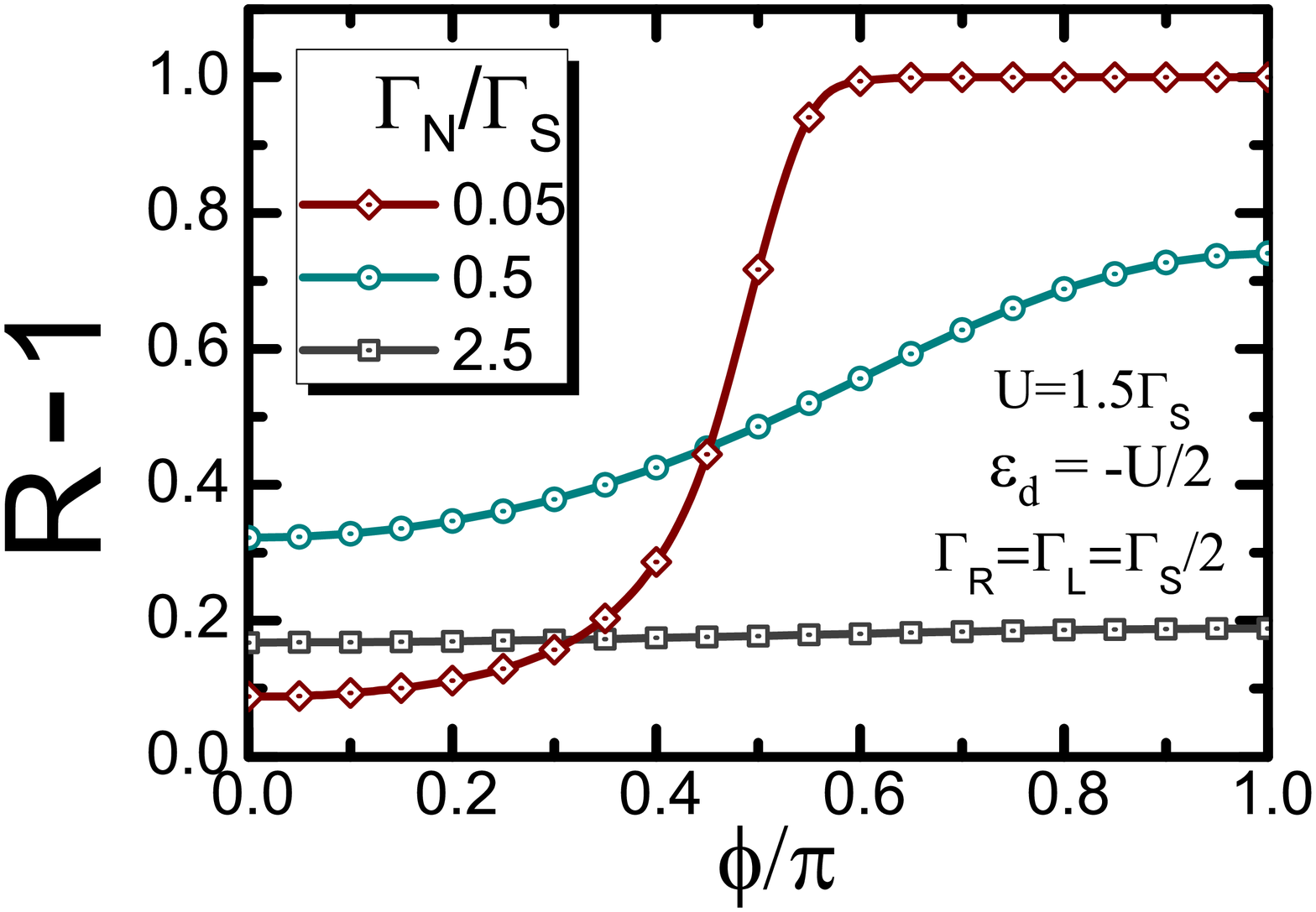}
\end{minipage}
\caption{
(Color online) 
Renormalized parameters are 
plotted vs Josephson phase $\phi$ 
 for several values of $\Gamma_N$ 
in the electron-hole symmetric case $\epsilon_d = -U/2$: 
 (upper panel) renormalized Andreev level $\pm \widetilde{E}_A$,
(lower panel) renormalization factor $z$ and  Wilson ratio $R$. 
The other parameters are chosen such that 
 $\Gamma_L=\Gamma_R$ ($= \Gamma_S/2$), 
and  $U=1.5\Gamma_S$ 
in the large gap limit $|\Delta_{L/R}| \to \infty$.  
}
\label{fig:Renorm_half}
\end{figure}

\subsubsection{Electron-hole symmetric case}
\label{sec:e-h_symmetric_case}

Figure  \ref{fig:YJ_cond_half} shows 
the results of the phase shift $\delta$, the SC pair correlation 
$2|\langle d_{\downarrow}^{} d_{\uparrow}^{} \rangle|$,  
the conductance $g_{NS}^{}$, and  
the Josephson current $J$ as functions of $\phi$ 
for several values of $\Gamma_N/\Gamma_S$ 
in the electron-hole symmetric case, 
where the Bogoliubov angle is 
locked at $\Theta_B=\pi/2$ and $\langle n_d \rangle =1$, 
as mentioned.
Thus, the ground state properties are determined 
by the phase shift $\delta$, which depends on 
the Josephson phase through  $E_A= \Gamma_S^{}\cos (\phi/2)$.
The Coulomb repulsion is chosen to be  $U=1.5\Gamma_S$,
and for this interaction the QPT occurs 
at $\phi \simeq 0.46 \pi$ in the $\Gamma_N \to 0$ limit. 

The coupling to the normal lead makes 
the excitation spectrum of the dot gapless and
changes the sharp QPT into a continuous crossover 
between the two different singlet states,  
namely the Kondo and local SC singlets. 
For $\phi \lesssim 0.46 \pi$,   
the ground state is the   
local Cooper pairing consisting of a linear combination of 
the empty and doubly occupied impurity states with small  $\delta$. 
For larger phase difference, $\phi \gtrsim 0.46 \pi$, however, 
the ground state is a Kondo singlet state, 
for which $\delta \simeq \pi/2$.

The phase shift shown in the upper left panel 
of Fig.\  \ref{fig:YJ_cond_half} can be expressed as 
the number of Bogoliubov particles on the impurity site, 
 $\langle n_{\gamma,-1} \rangle =2\delta/\pi$, 
due to the Friedel sum rule given in Eq.\ \eqref{MBogo}.
Furthermore in the electron-hole symmetric case, 
the SC pair correlation defined in Eq.\ \eqref{eq:K_sDQD} 
is given simply by  
 $2\langle  d_{\uparrow}^{} d_{\downarrow}^{}\rangle 
=1-\langle n_{\gamma,-1} \rangle$.
Therefore, the pair correlation is suppressed in the Kondo regime for 
$\phi \gtrsim 0.46 \pi$, 
as seen in the upper right panel.

The Andreev conductance $g_{NS}^{}$, 
shown in the lower left panel 
of Fig.\  \ref{fig:YJ_cond_half},
also varies as a function of the Josephson phase $\phi$. 
At the crossover region between the Kondo singlet 
and local-Cooper-pairing states, 
the conductance due to the Andreev scattering $g_{NS}^{}$ has 
a sharp peak for small $\Gamma_N^{}$.
The crossover behavior, however, is smeared as $\Gamma_N$ increases.
The conductance takes the unitary limit value $4e^2/h$ at $\delta=\pi/4$ 
where the renormalized resonance width 
and the renormalized Andreev level  coincide  
such that $\widetilde{\Gamma}_N^{}=\widetilde{E}_A^{}$.
This happens in Fig.\ \ref{fig:YJ_cond_half} 
at $\phi \simeq 0.0$ for $\Gamma_N^{}=0.5\Gamma_S$.

The Josephson current, in the lower right panel, 
also shows the crossover behavior near $\phi \simeq 0.46 \pi$,
and decreases at $\phi \gtrsim 0.46 \pi$ when $\Gamma_N$ is small. 
The value of the current approaches 
zero in the Kondo-singlet region since in this case 
the large gap limit has been taken.
For finite SC gaps, however, a weak current will    
flow in the opposite direction  
as seen in Fig.\ \ref{Fig:results_2SC_J_dd} 
for the magnetic-doublet state. 
Similarly,  for finite SC gaps,
the SC pair correlation 
 $\langle  d_{\uparrow}^{} d_{\downarrow}^{}\rangle$ 
will also change the sign at crossover, 
and has a small negative value in the magnetic ground state.

We have also deduced the renormalized parameters for 
the local Fermi liquid of  the interacting Bogoliubov particles 
from the convergence of the finite-size energy spectrum 
in the successive NRG steps.\cite{KWW2,Hewson2}
Figure \ref{fig:Renorm_half} shows the results 
for the renormalized Andreev level 
 $\pm \widetilde{E}_A^{}$, the wavefunction renormalization factor $z$, 
and the Wilson ratio $R$.
We see for small $\Gamma_N$ ($=0.05\Gamma_S$) 
that the parameters are significantly renormalized 
in the Kondo singlet state for $\phi \gtrsim 0.46 \pi$, 
where $z \ll 1.0$,  $R \simeq 2.0$, and 
the pair of  renormalized Andreev 
peaks $\pm \widetilde{E}_A$ lie 
close to the Fermi level $\omega=0$.  
This indicates that the Bogoliubov particles 
are strongly correlated in the Kondo regime. 
In contrast,  
 in the local Cooper-pairing state for $\phi \lesssim 0.46 \pi$, 
the parameters are not renormalized so much 
 $z \simeq 1.0$, $R \simeq 1.1$, and 
the renormalized Andreev peaks  $\pm \widetilde{E}_A^{}$ situate 
away from the Fermi level.   
As the coupling 
$\Gamma_N$ between the QD and normal lead becomes large,
these two singlet states  become indistinguishable, 
as those for $\Gamma_N=2.5 \Gamma_S$ 
seen in Fig.\ \ref{fig:Renorm_half}.

\begin{figure}[t]

\begin{minipage}{1\linewidth}
 \leavevmode
\includegraphics[width=0.48\linewidth]{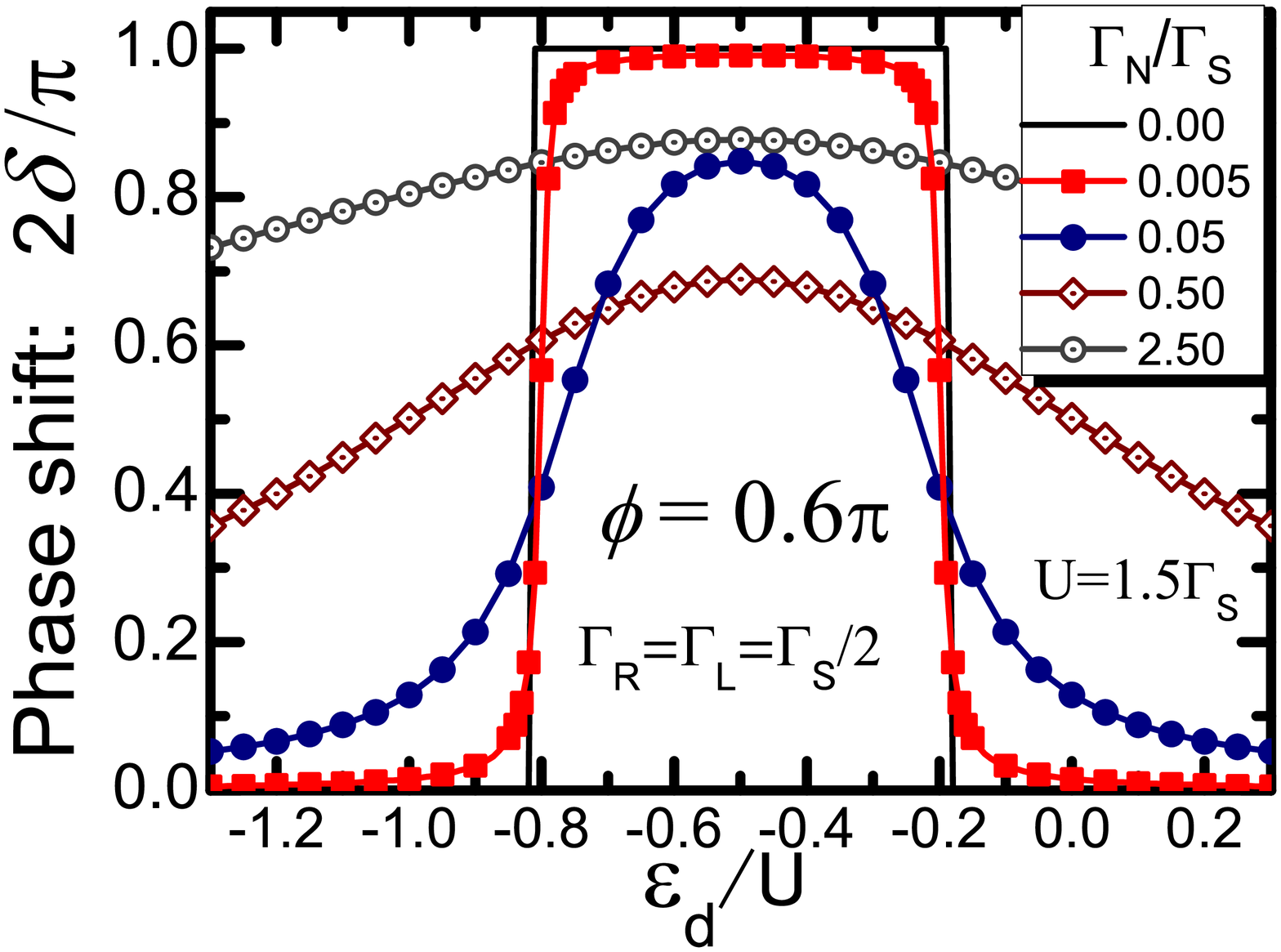}
\rule{0.01\linewidth}{0cm}
\includegraphics[width=0.48\linewidth]{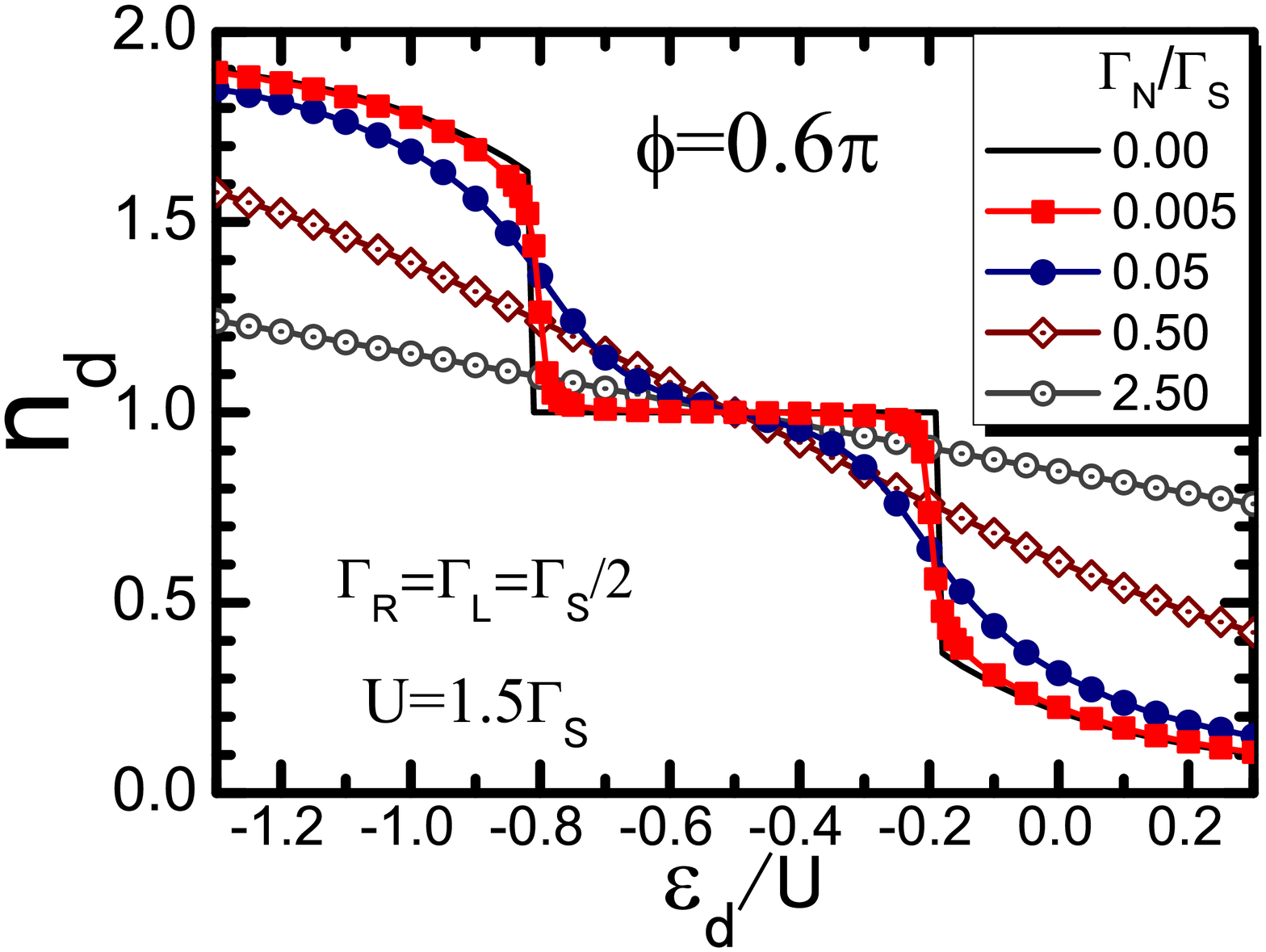}
\includegraphics[width=0.48\linewidth]{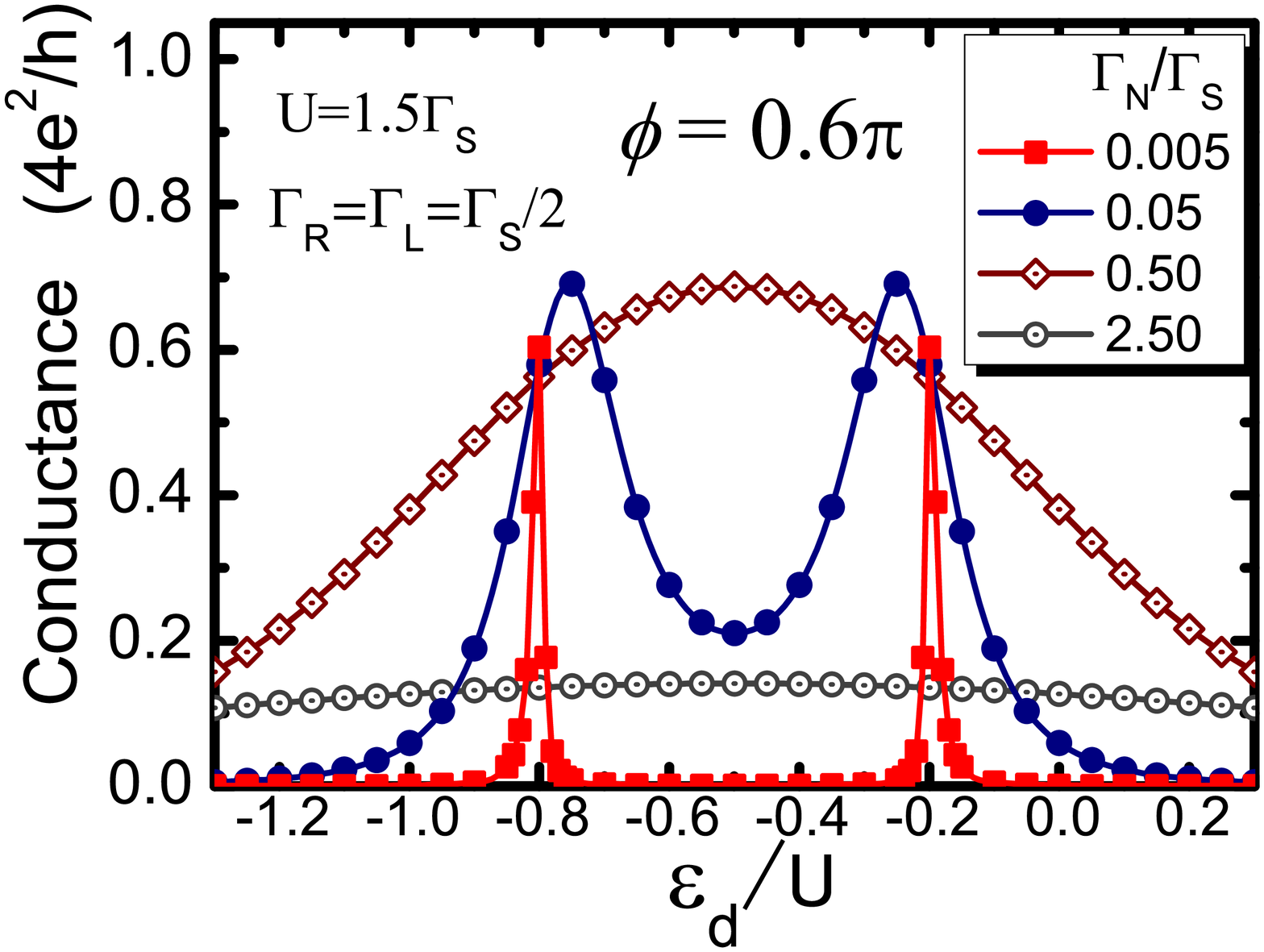}
\rule{0.01\linewidth}{0cm}
\includegraphics[width=0.485\linewidth]{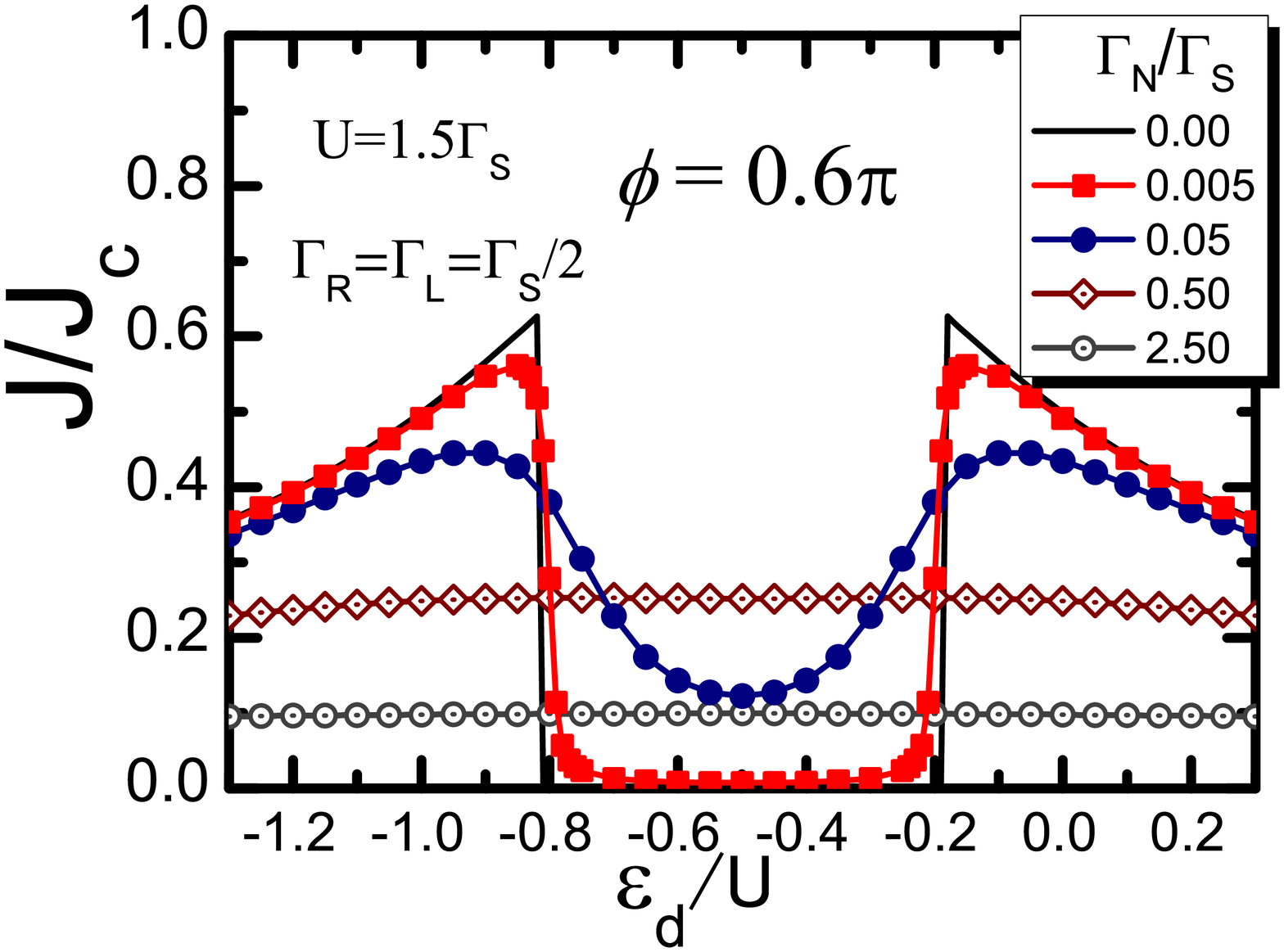}
\end{minipage}
\caption{
(Color online)  
Phase shift and some related ground-state averages   
are plotted vs $\epsilon_d/U$ 
 for $\phi=0.6\pi$ for several values of $\Gamma_N$:
(upper panel) phase shift $\delta$ and
 electron occupation number $\langle n_d \rangle$, 
(lower panel) conductance $g_{NS}^{}$ 
and Josephson current $J$ in units of $J_C = e\Gamma_S^{}/\hbar$.
The parameters are chosen to be  
 $\Gamma_L=\Gamma_R$ ($= \Gamma_S/2$), 
and  $U=1.5\Gamma_S$. 
}
\label{fig:YJ_cond_phi06}
\end{figure}

%

\subsubsection{Electron-hole asymmetric case}
\label{sec:e-h_asymmetric_case}

We next consider the electron-hole asymmetric case, in which 
the Bogoliubov angle $\Theta_B=\cot^{-1}(\xi_d/|\Delta_d|)$ deviates 
from $\pi/2$  as $\xi_d$ varies.
Furthermore, the phase shift $\delta$ also varies as a 
a function of $\xi_d$  
since the bare position $E_A-U/2$ of the Andreev level depends 
on $\xi_d$. 
Through these changes of the phase parameters $\Theta_B$ and $\delta$, 
the ground-state properties of this Y-junction  
depend on the gate voltage $\epsilon_d$.

In this subsection we examine the $\epsilon_d$ dependence 
for several values of the 
Josephson phase $\phi=0.3\pi,\, 0.46\pi$, and $0.6\pi$, 
choosing the Coulomb repulsion to be  $U=1.5\Gamma_S$ 
as that in the half-filled case discussed in the above. 
The phase boundary between the singlet and doublet ground states 
moves in the  $\Gamma_S^{}$ vs $\epsilon_d$ plane as $\phi$ increases 
as shown in Fig.\ \ref{fig:phase_diagram_atomic_limit}. 
In our parameter set  $\Gamma_S/U = 0.666$, 
and the QPT occurs for $\phi=0.6\pi$  
when $\epsilon_d$ varies in the range  $ -1.0 \leq \epsilon_d/U  \leq 0.0$,   
whereas the level crossing does not occur for $\phi=0.3\pi$.
A marginal situation is realized for $\phi\simeq 0.46\pi$, 
in this case the system approaches very 
closely to the phase boundary 
near the symmetric point $\epsilon_d\simeq -U/2$.

\begin{figure}[t]

\begin{minipage}{1\linewidth}
 \leavevmode
\includegraphics[width=0.48\linewidth]{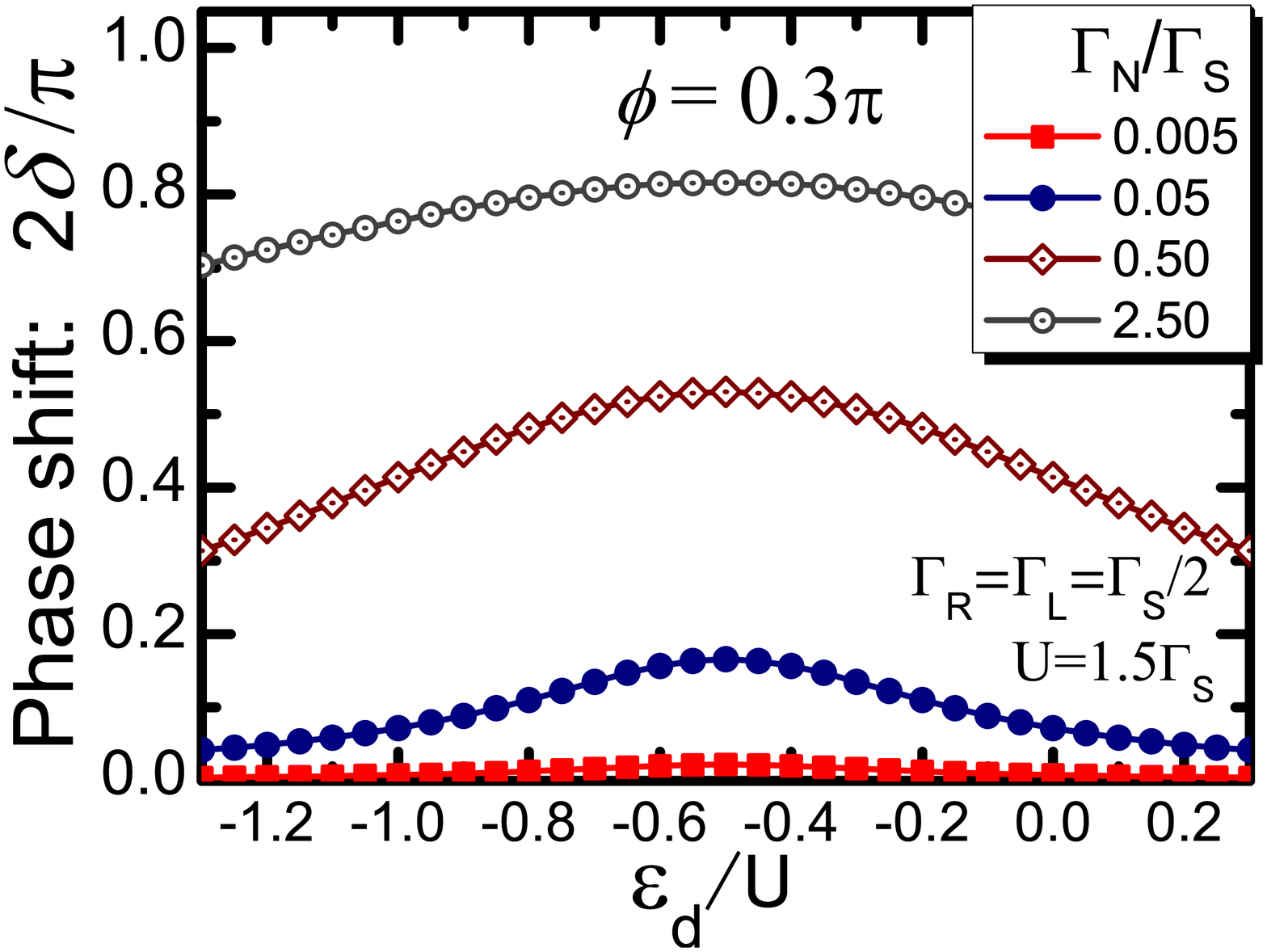}
\rule{0.01\linewidth}{0cm}
\includegraphics[width=0.48\linewidth]{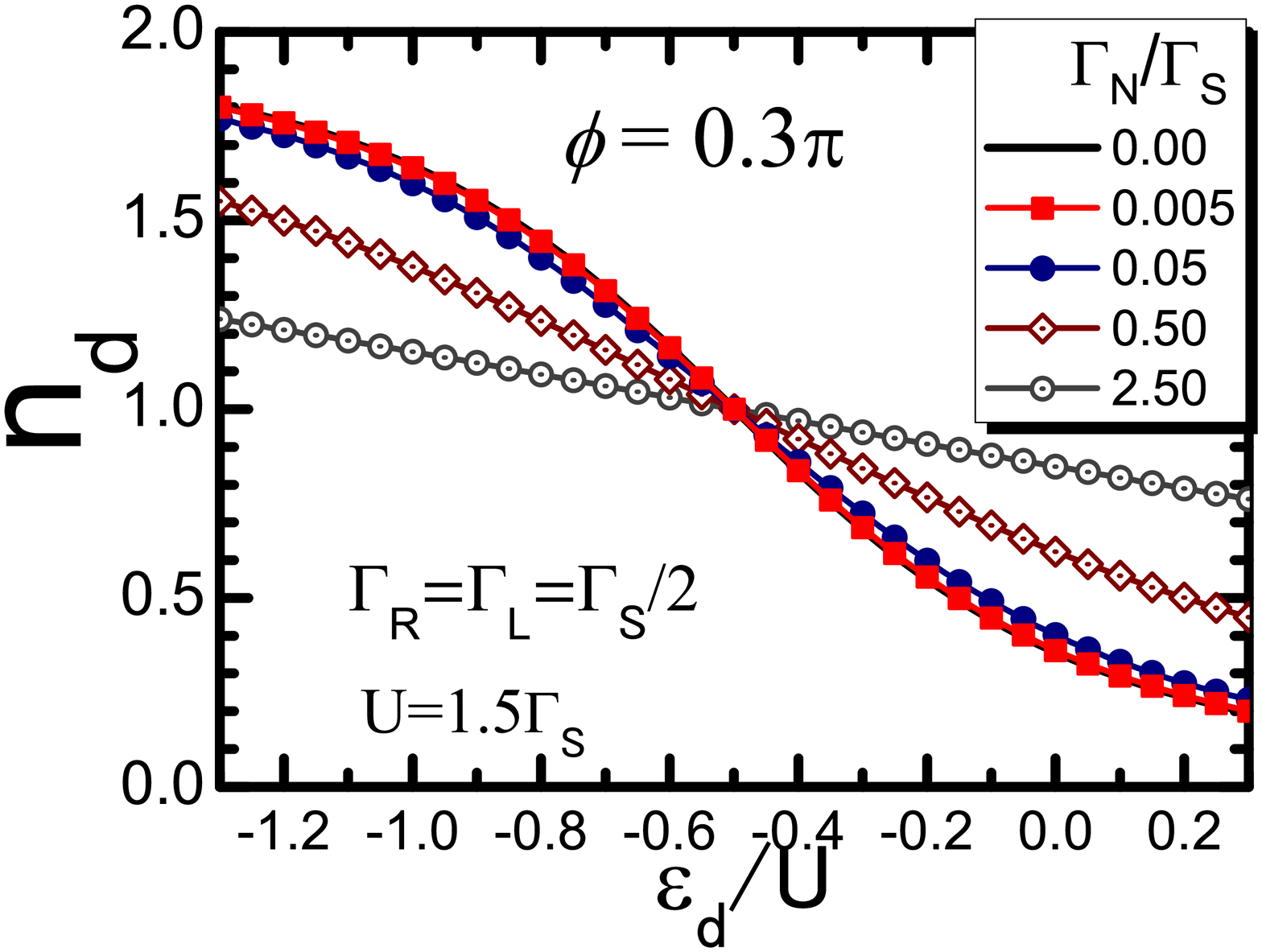}
\includegraphics[width=0.48\linewidth]{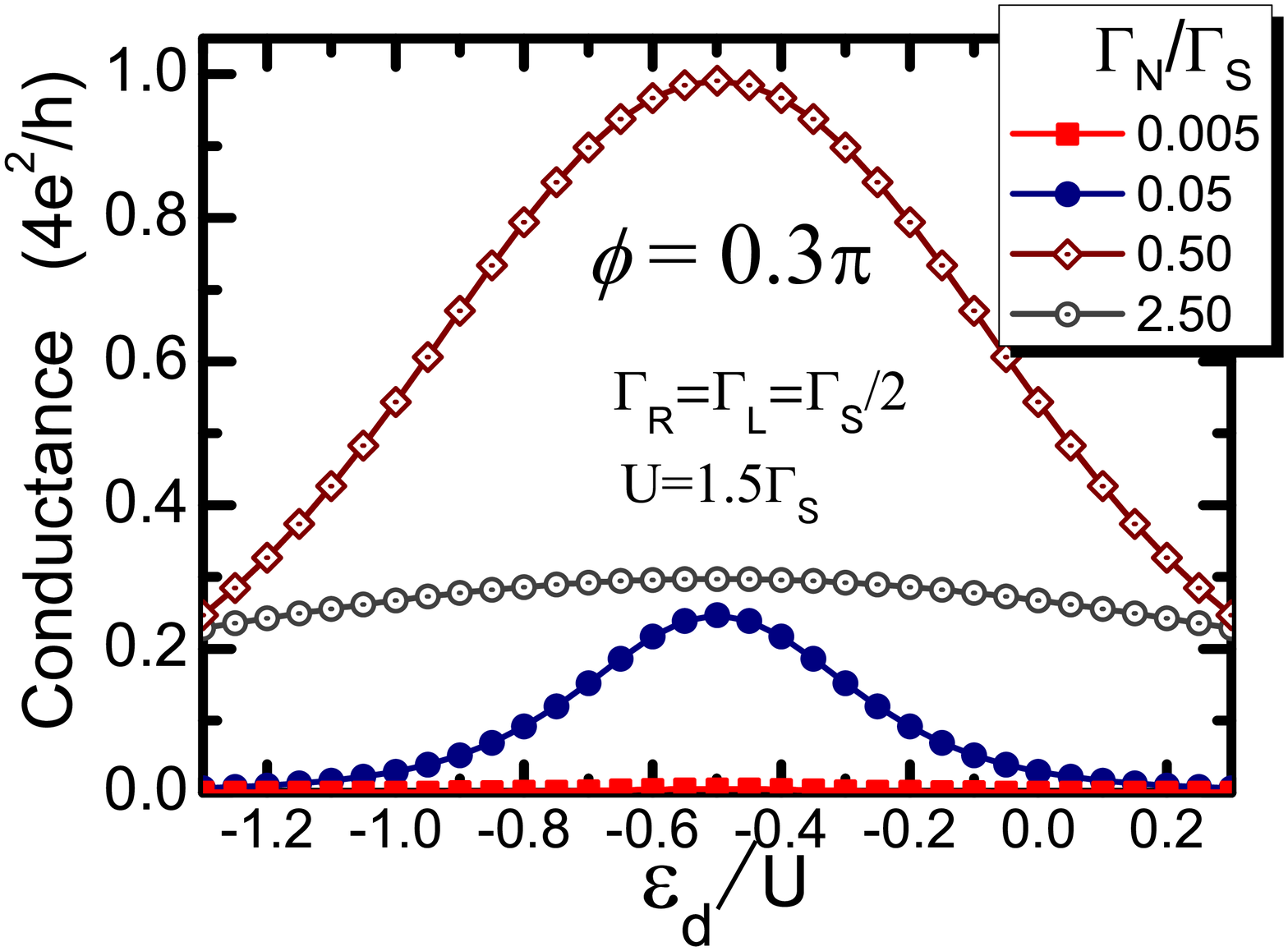}
\rule{0.01\linewidth}{0cm}
\includegraphics[width=0.48\linewidth]{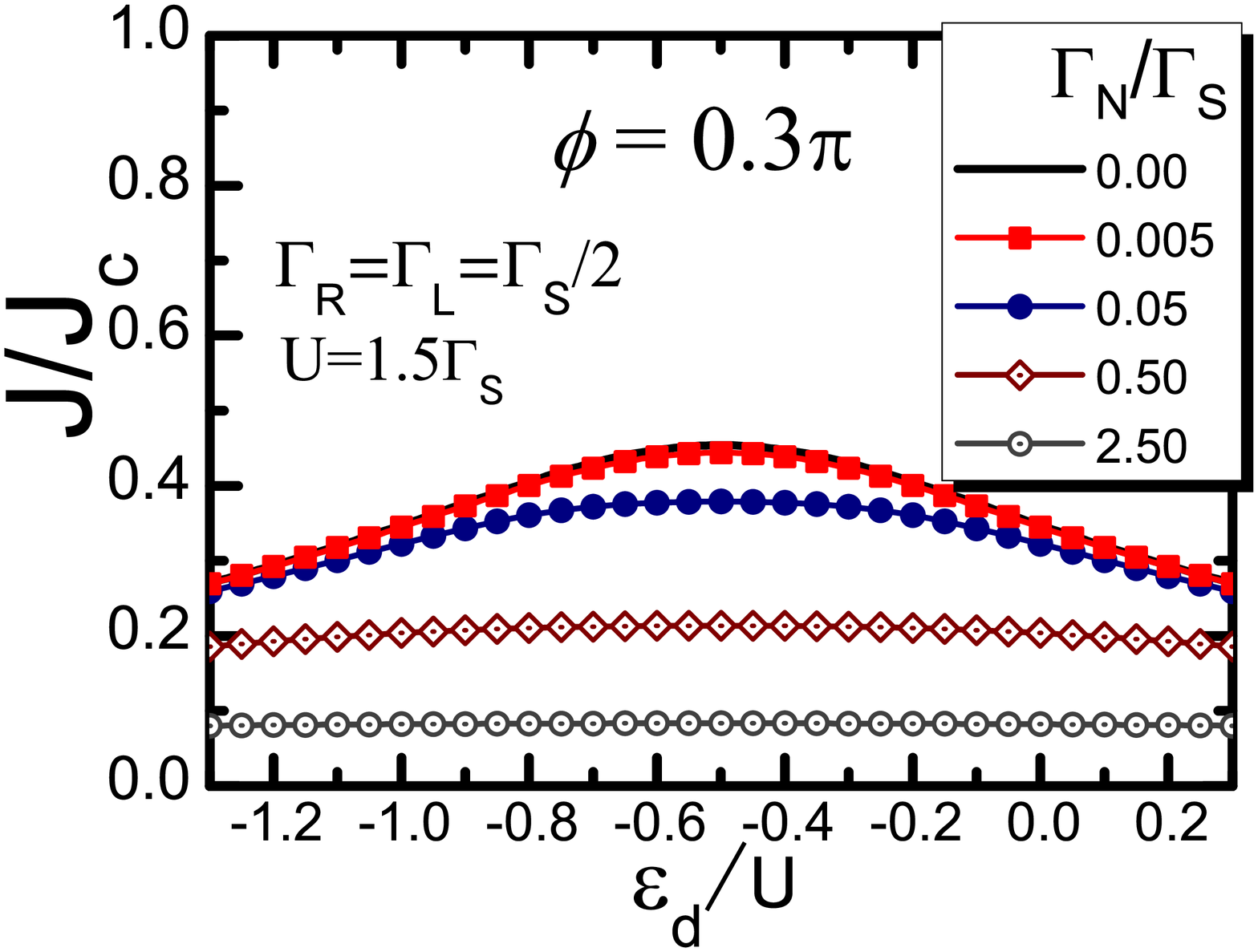}
\end{minipage}
\caption{
(Color online) 
Phase shift and some related ground-state averages   
are plotted vs $\epsilon_d/U$ 
 for $\phi=0.3\pi$ for several values of $\Gamma_N$:
(upper panel) phase shift $\delta$ and
 electron occupation number $\langle n_d \rangle$, 
(lower panel) conductance $g_{NS}^{}$ 
and Josephson current $J$ in units of $J_C = e\Gamma_S^{}/\hbar$.
The parameters are chosen to be  
 $\Gamma_L=\Gamma_R$ ($= \Gamma_S/2$), 
and  $U=1.5\Gamma_S$. 
}
\label{fig:YJ_cond_phi03}
\end{figure}

In Fig.\ \ref{fig:YJ_cond_phi06},  
the correlation functions for $\phi=0.6\pi$ 
are plotted vs $\epsilon_d/U$.  
The ground state changes discontinuously in the limit of $\Gamma_N^{} = 0$,  
at $\epsilon_d \simeq -0.82U$ and $-0.18U$. 
The sharp transition becomes a continuous crossover for finite $\Gamma_N^{}$,
and at $ -0.82U \lesssim \epsilon_d \lesssim -0.18U$ the ground state 
is a Kondo singlet consisting of the strong correlated Bogoliubov particles, 
and the electron filling is almost $\langle n_d \rangle \simeq 1.0$, 
as shown in the upper right panel.
In the upper left panel, we can see that in this region of $\epsilon_d$  
the occupation number of the Bogoliubov particles 
at the impurity level, $\langle n_{\gamma,-1}^{} \rangle = 2 \delta/\pi$, 
decreases as  
 $\Gamma_N^{}$ increases from zero to an intermediate 
value $\Gamma_N^{} \lesssim 0.5 \Gamma_S^{}$. 
Then,  $\langle n_{\gamma,-1}^{}\rangle$  increases as seen 
in the results obtained for $\Gamma_N^{}=2.5\Gamma_S^{}$,  
and approaches $1.0$ in the limit of large $\Gamma_N^{}$ where 
this coupling dominates all the other effects. 
The lower right panel of Fig.\ \ref{fig:YJ_cond_phi06}  
shows the Josephson current.
This supercurrent, flowing between the two SC leads,  
is suppressed due to the electron correlation in the Kondo regime 
at $ -0.82U \lesssim \epsilon_d \lesssim -0.18U$. 
Outside of this region 
the ground state is characterized 
by the local Cooper pairing, and 
the current is less suppressed  although 
the coupling to the normal lead $\Gamma_N^{}$
smears the structure due to the QPT.
We also see in the lower left panel that 
the conductance $g_{NS}^{}$ between the normal lead and the QD 
shows sharp peak
at the transient region of the crossover  for small $\Gamma_N^{}$. 
The sharp conductance peak is mainly caused 
by the phase shift $\delta$ that 
changes suddenly from $0$ to $\pi/2$
at the crossover region because  
the conductance is proportional to $\sin^2 2\delta$.
The Bogoliubov angle  $\Theta_B$, 
appearing in the expression of  $g_{NS}^{}$   
given in Eq.\ \eqref{eq:Cond}, varies moderately and determines 
the peak height.

Figure \ref{fig:YJ_cond_phi03} shows 
the $\epsilon_d$ dependence of 
the correlation functions for a smaller value of the 
Josephson phase $\phi=0.3\pi$.
In this case  $U$ is not large enough to 
reach the Kondo regime over the crossover region. 
The ground state is the local Cooper-pairing state 
for all values of $\epsilon_d$, and thus  
the correlation functions vary moderately 
as a function of the gate voltage $\epsilon_d$.
The conductance  and Josephson current
have maximum at the electron-hole symmetric point 
$\epsilon_d = -0.5U$. 
This is mainly because the factor $\sin \Theta_B$ that 
appears in the expression of these correlations given in 
Eqs.\ \eqref{eq:Cond} and \eqref{eq:JosephsonCurrentY-junction}
takes a local maximum at the Bogoliubov angle of $\Theta_B=\pi/2$.

\begin{figure}[t]

\begin{minipage}{1\linewidth}
 \leavevmode
\includegraphics[width=0.5\linewidth]{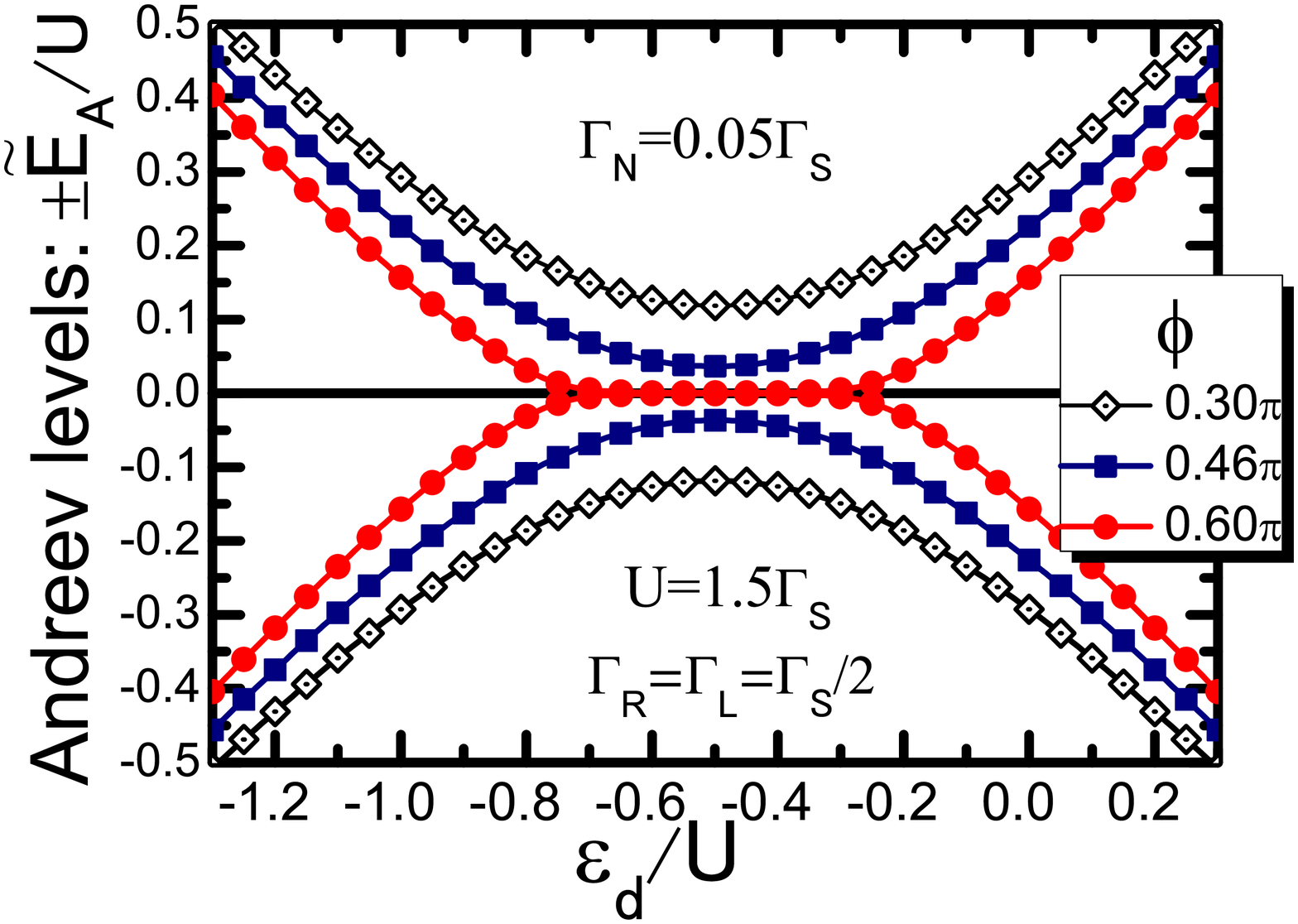}
\includegraphics[width=0.475\linewidth]{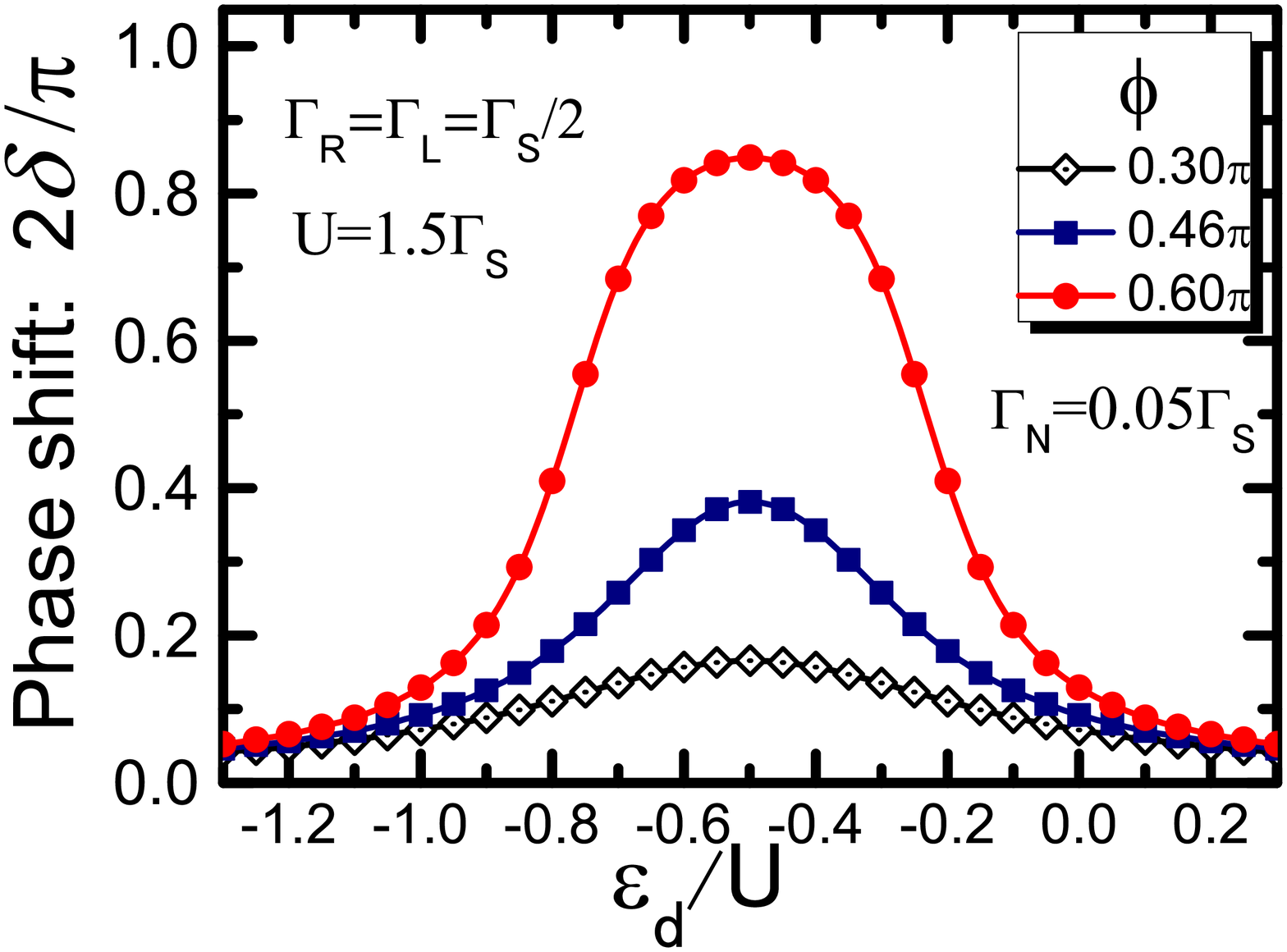}
\rule{0.01\linewidth}{0cm}
\includegraphics[width=0.46\linewidth]{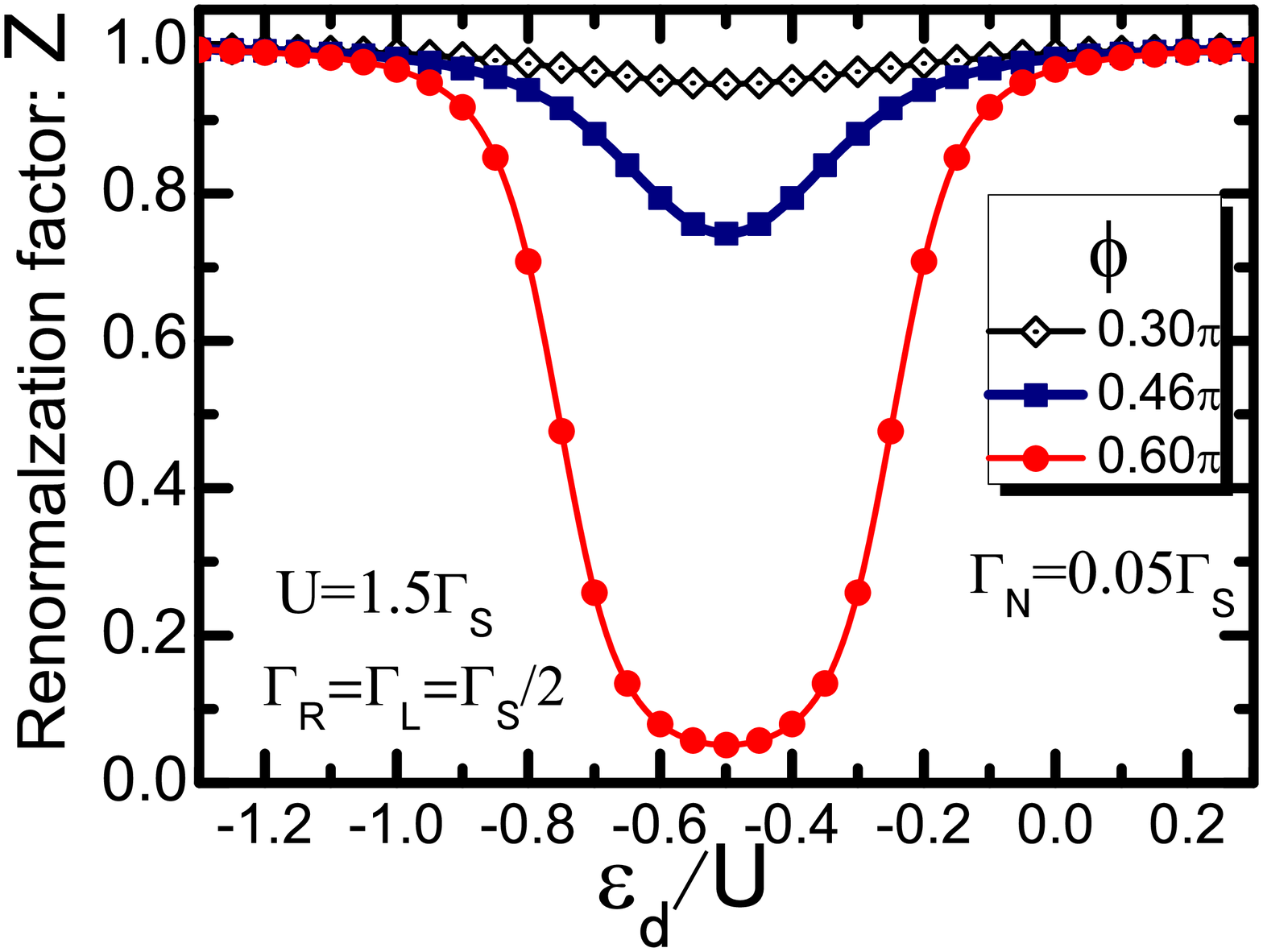}
\includegraphics[width=0.48\linewidth]{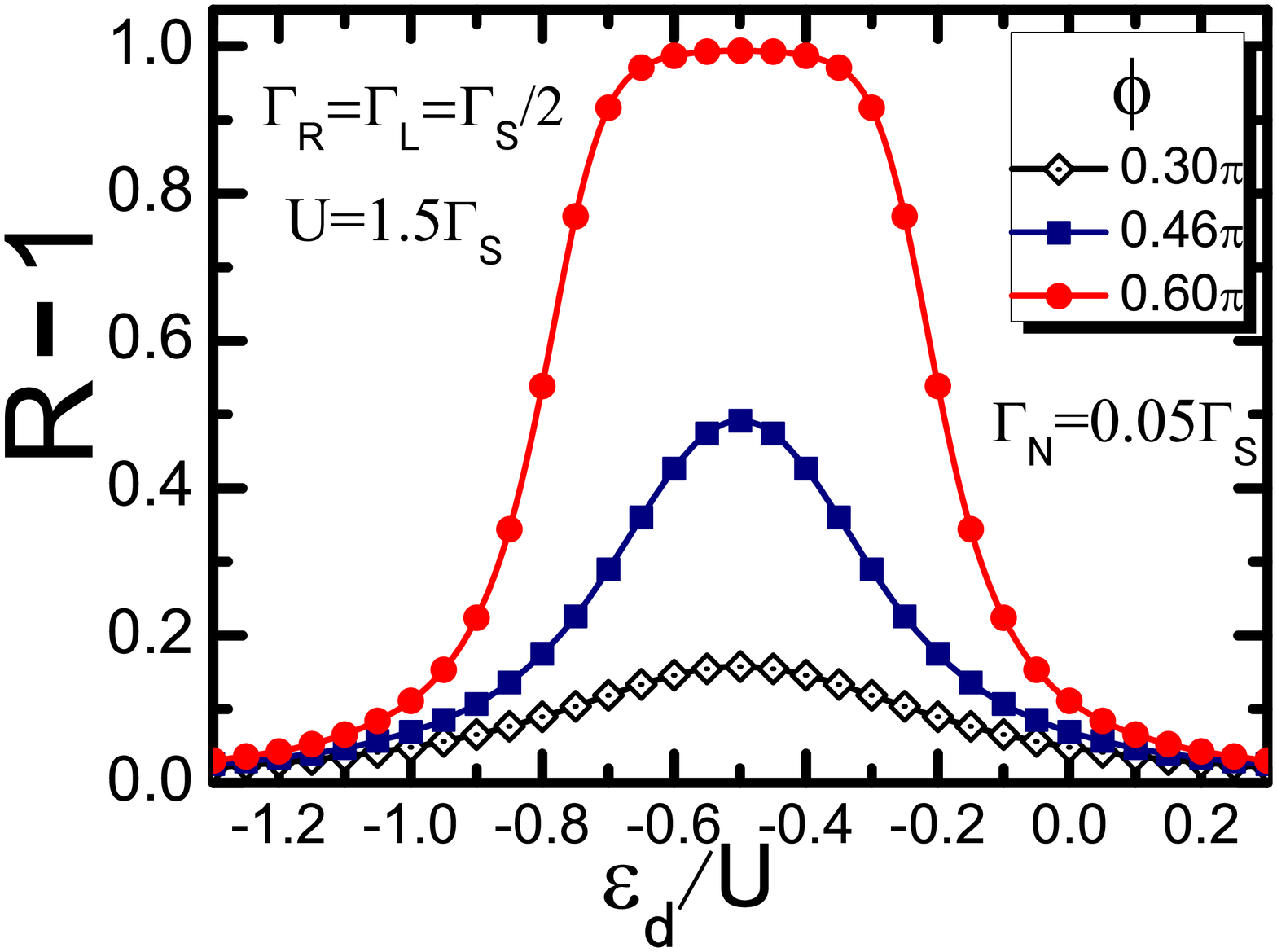}
\end{minipage}
\caption{
(Color online)
Renormalized parameters are 
plotted vs $\epsilon_d/U$ for several values of $\phi$  
for relatively small coupling $\Gamma_N=0.05\Gamma_S$ 
between the dot and normal lead: 
 (upper panel) 
renormalized Andreev level $\pm \widetilde{E}_A$ and phase shift $\delta$, 
(lower panel) renormalization factor $z$ and  Wilson ratio $R$. 
The other parameters are chosen such that 
 $\Gamma_L=\Gamma_R$ ($= \Gamma_S/2$), and  $U=1.5\Gamma_S$. 
}
\label{fig:Renorm_ed_some_ph}
\end{figure}


In Fig.\ \ref{fig:Renorm_ed_some_ph}, 
the results of the renormalization factor $z$, 
the renormalized Andreev levels $\pm \widetilde{E}_A$, 
the Wilson ratio $R$, 
and the phase shift are compared for three different values for the 
Josephson phase $\phi =0.3\pi,\, 0.46\pi,$ 
and $0.6\pi$. 
The Coulomb interaction and the hybridization energy scales 
are chosen such that $U=1.5\Gamma_S$ and $\Gamma_N=0.05\Gamma_S$.
We can see in the upper left panel that 
the pair of renormalized Andreev levels $\pm \widetilde{E}_A$ 
for $\phi=0.6 \pi$ lie very closely to the Fermi level 
at $ -0.82U \lesssim \epsilon_d \lesssim -0.18U$.
Furthermore, 
in this region, the renormalization factor $z$  
significantly decreases and the Wilson ratio 
approaches $R\to 2$ 
owing to the strong correlations in the Kondo regime.
It also indicates that the Kondo temperature 
$T_K=\pi\widetilde{\Gamma}_N/4$ and the renormalized 
resonance width $\widetilde{\Gamma}_N$, 
defined in Eq.\ \eqref{eq:z_factor},
become very small. 
Simultaneously, the local Bogoliubov particle number on the dot approaches  
single occupancy  $\langle n _{\gamma,-1} \rangle
=2\delta/\pi \simeq 0.9$ although it is less than $1.0$ 
because $\Gamma_N^{}$ is not very small in this case. 
Note that the sharp single Kondo peak that 
we have seen in Fig.\ \ref{fig:spec_phi=0} 
for $U\gtrsim 15 \Gamma_N$ consists of such 
a pair of  renormalized Andreev levels, 
appearing in the close vicinity of the Fermi level.

The ground-state properties show a marginal behavior 
at $\phi=0.46\pi$, as shown in Fig.\ \ref{fig:Renorm_ed_some_ph}.
For instance, $z$ and $R$ take 
the intermediate values, 
and the pair of $\pm \widetilde{E}_A$ becomes  distinguishable 
as we can see in the upper left panel of 
Fig.\ \ref{fig:Renorm_ed_some_ph}.
Then, for a smaller value of the Josephson phase $\phi=0.3\pi$, 
the ground state is a singlet caused by the local 
Cooper pairing, as mentioned. Therefore, 
in this case the electron correlations are suppressed as 
$z\simeq 0.95$ and  $R\simeq 1.16$ even at the 
electron-hole symmetric point $\epsilon_d=-0.5U$.

\begin{figure}[t]

\begin{minipage}{1\linewidth}
 \leavevmode
\includegraphics[width=0.483\linewidth]{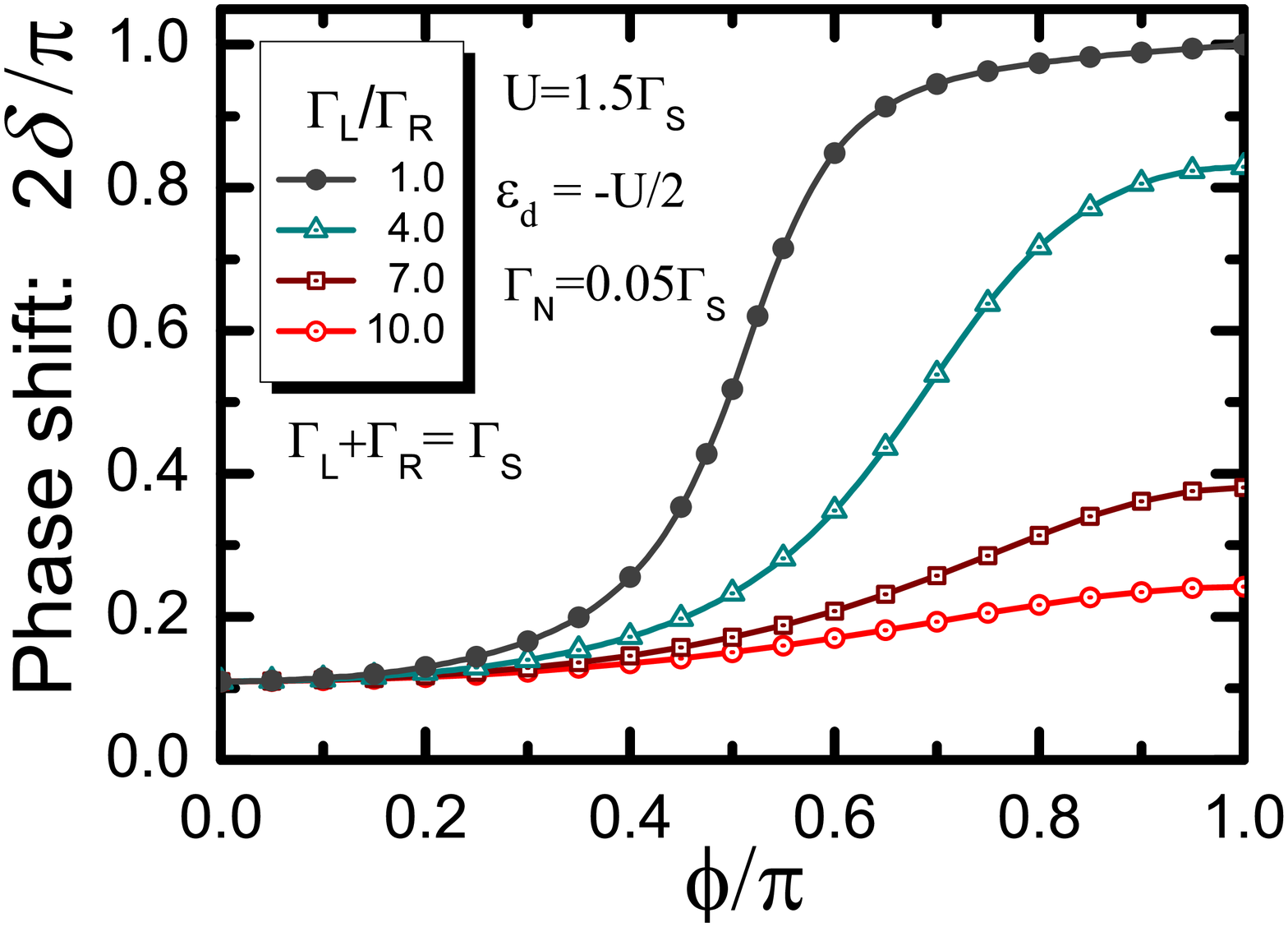}
\rule{0.01\linewidth}{0cm}
\includegraphics[width=0.483\linewidth]{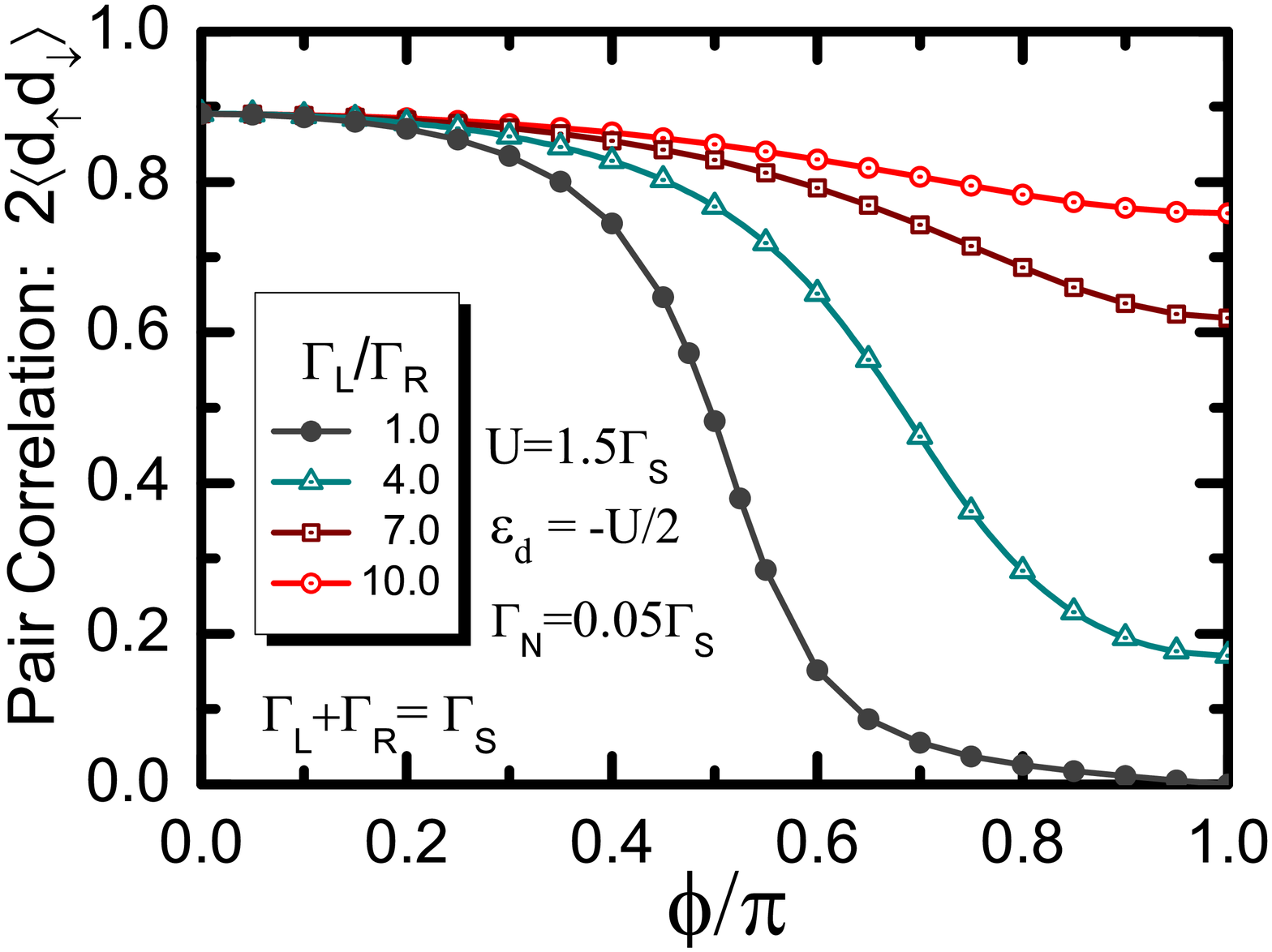}
\includegraphics[width=0.483\linewidth]{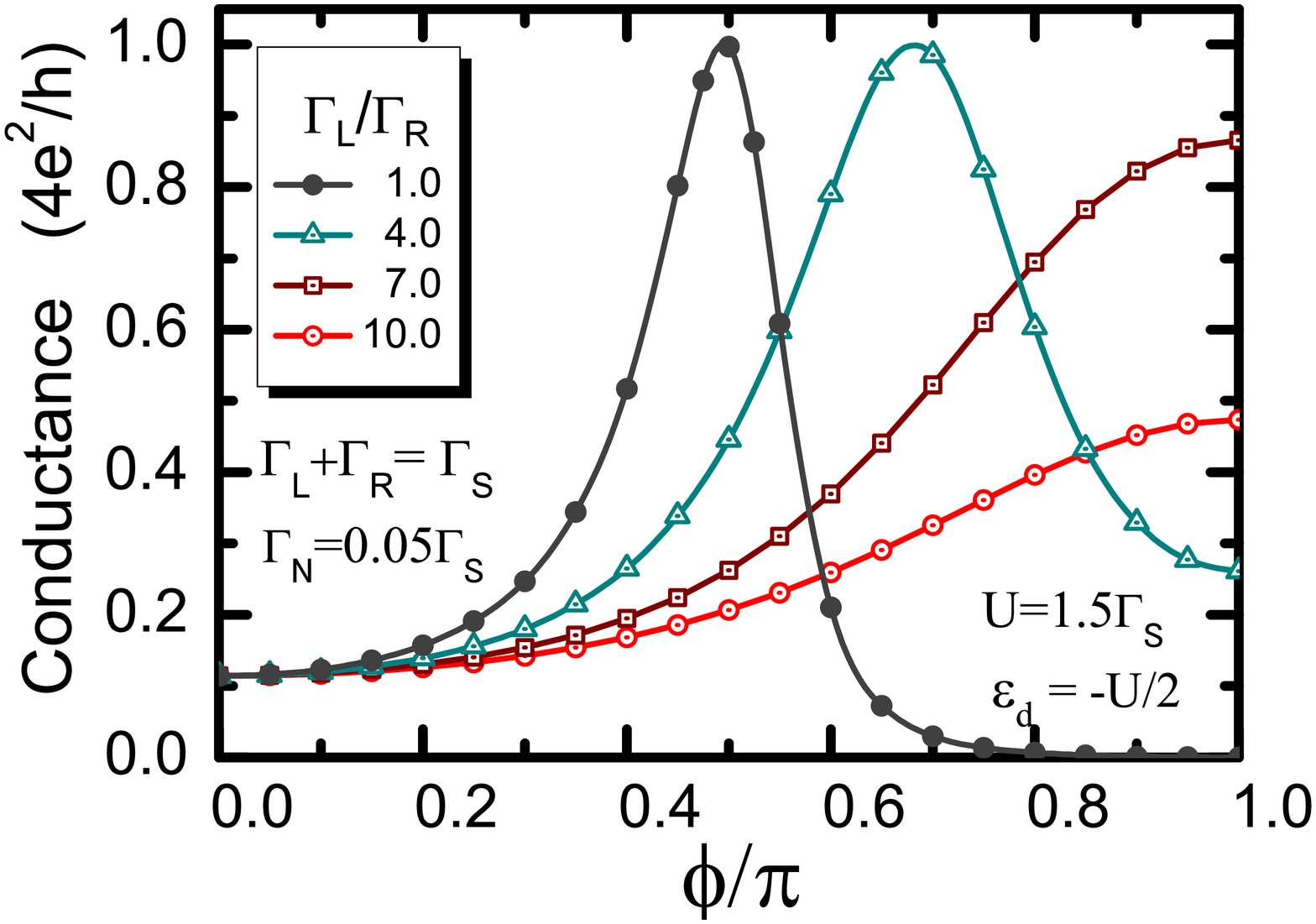}
\rule{0.01\linewidth}{0cm}
\includegraphics[width=0.483\linewidth]{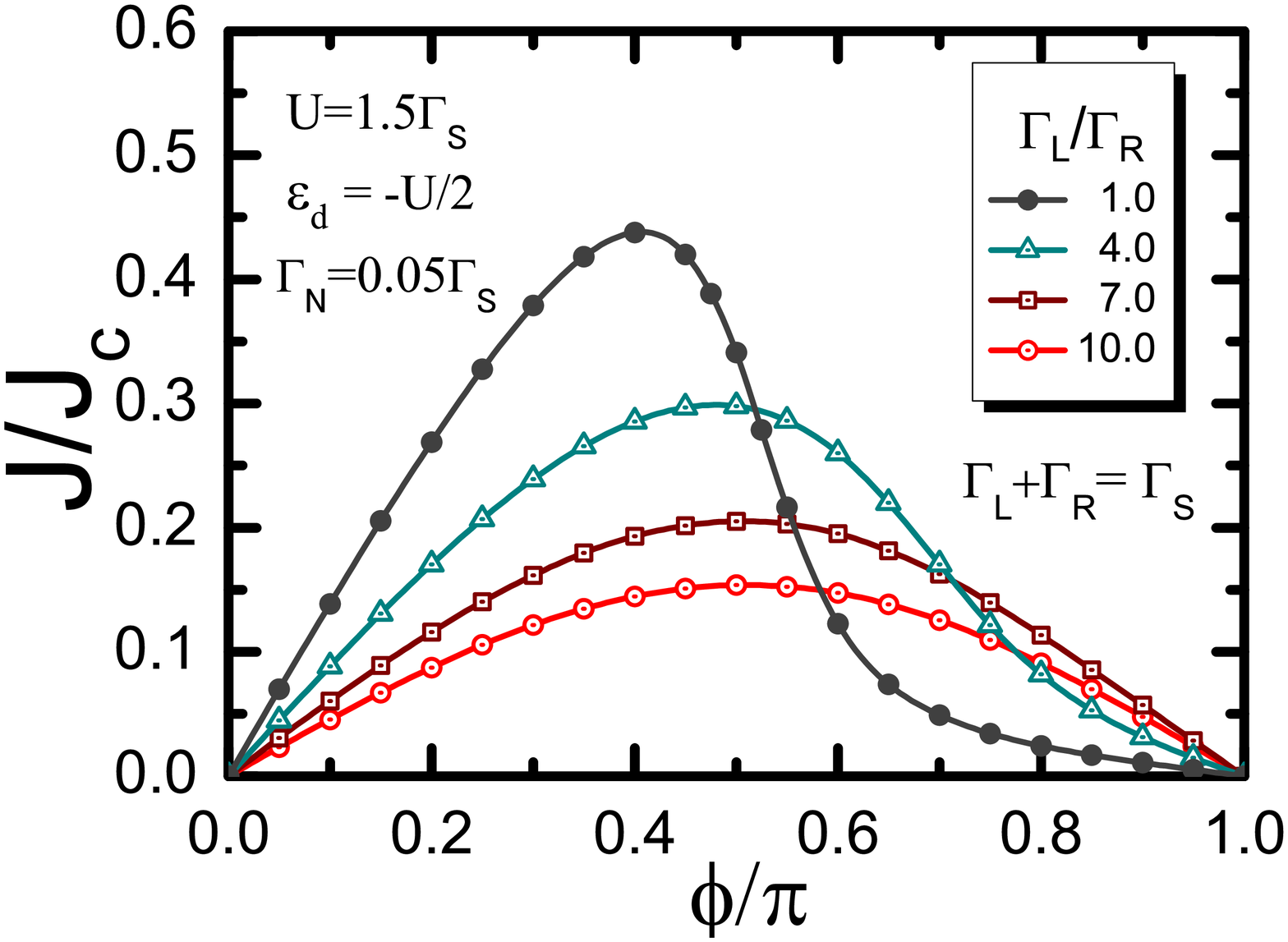}
\end{minipage}
\caption{
(Color online) 
Phase shift and some related ground-state averages
for the asymmetric couplings $\Gamma_R \neq \Gamma_L$ 
are plotted vs  Josephson phase $\phi$ 
for several $\Gamma_L/\Gamma_R$ keeping 
$\Gamma_S =\Gamma_R+\Gamma_L$ unchanged:  
(upper panel) phase shift $\delta$ and pair correlation  
$2\left|\langle d_{\downarrow}^{} d_{\uparrow}^{} \rangle\right|$, 
(lower panel) conductance $g_{NS}^{}$ and 
(lower right) Josephson current in units of $J_C = e\Gamma_S^{}/\hbar$.
The other parameters are chosen such that $\epsilon_d = -U/2$, 
 $U=1.5\Gamma_S$ and $\Gamma_N=0.05\Gamma_S$. 
}
\label{fig:YJ_cond_asm}
\end{figure}

%


\subsection{Spatial asymmetry in the junction
 ($\Gamma_L \neq \Gamma_R$)}
\label{subsec:asymmetric_junction}

So far, we have assumed that the Josephson junction 
is symmetric $\Gamma_L = \Gamma_R$. 
However, 
the SC proximity effects in real systems 
depend on the asymmetry 
in the couplings $\Gamma_L \neq \Gamma_R$.
One difference of the asymmetric junction 
from the symmetric one 
is that the maximum possible of the transmission probability 
is no longer one, namely  $\mathcal{T}_0 < 1$. 
Specifically, the local SC gap induced on the dot 
$|\Delta_d|$,
defined in Eq.\ \eqref{eq:Delta_d_def},  
becomes larger in the asymmetric junction than that in  
the symmetric junction for which $\mathcal{T}_0=1$.
Therefore, the asymmetry in the couplings 
tends to enhance the SC proximity effects 
as it suppresses the reduction 
of $|\Delta_d|$ due to the Josephson phase $\phi$.
In the following, we discuss the ground state properties 
of the asymmetric junction with $\Gamma_L \neq \Gamma_R$, 
keeping the sum $\Gamma_S=\Gamma_R+\Gamma_L$ 
unchanged at $\Gamma_N=0.05\Gamma_S$. 
For simplicity, 
we examine the electron-hole symmetric case $\epsilon_d=-U/2$, 
where the bare Andreev level is given by 
$E_A =|\Delta_d|$ 
and the Bogoliubov angle is fixed at $\Theta_B=\pi/2$.

The NRG results for the asymmetric junction  
are plotted vs the Josephson phase $\phi$ 
in Fig.\ \ref{fig:YJ_cond_asm} for $U=1.5\Gamma_S$. 
In each panel, the curve for $\Gamma_L/\Gamma_R=1.0$ corresponds to the 
results obtained for the symmetric coupling, 
presented also in Fig.\ \ref{fig:YJ_cond_half}.
As mentioned in the above, 
the amplitude of the static SC gap $|\Delta_d|$ 
for the asymmetric junction 
becomes larger that for the symmetric junction.
Thus the position of $E_A$ moves away from the Fermi energy 
due to the spatial asymmetry, 
and this causes rather moderate $\phi$ dependence of 
the phase shift $\delta$ for $\Gamma_L \neq \Gamma_R$, 
seen in the upper left panel of Fig.\ \ref{fig:YJ_cond_asm}.
Correspondingly, the SC pair correlation 
$2\langle d_{\uparrow}^{} d_{\downarrow}^{}  \rangle$, 
shown in the upper right panel, 
increases as the asymmetry  $\Gamma_L/\Gamma_R$ increases.

The conductance due to the Andreev scattering  
is proportional to $\sin^2 2 \delta$, 
and thus has a peak when the phase shift 
takes the value of $\delta=\pi/4$ 
at crossover region between the local-Cooper pairing 
and Kondo singlet states.   
We see in the lower left panel 
of Fig.\ \ref{fig:YJ_cond_asm} 
that the position of the conductance peak shifts 
towards the larger $\phi$ side   
as $\delta$ becomes smaller 
with increasing $\Gamma_L/\Gamma_R$.
This also shows that 
the asymmetric coupling favors the SC proximity into the dot,
and enlarges the parameter region  
for the local-Cooper-pairing ground state. 
Furthermore, 
the crossover behavior from the local-Cooper pairing to the Kondo regime, 
seen for the Josephson current in the lower right panel 
for $\Gamma_L/\Gamma_R=1$,
is smeared as $\Gamma_L/\Gamma_R$ increases, 
and the current shows a simple sinusoidal $\phi$ dependence 
for large asymmetries.


\section{Summary}
\label{sec:Summary}

We have studied the crossover between 
a Kondo singlet and a local-Copper-pairing singlet,   
occurring in a quantum dot coupled to one normal and two SC leads. 
The low-energy states of the system can be described 
by excitations from a local Fermi-liquid ground state 
of interacting Bogoliubov particles.
Specifically in this three terminal configuration, 
the renormalized parameters for the quasiparticles  
vary as functions of the Josephson phase $\phi$, and  
the crossover occurs at finite $\phi_C^{}$ 
as the  phase difference varies in the range $0\leq |\phi| \leq \pi$.
We have calculated the phase shift $\delta$,  
the renormalization factor $z$, the renormalized 
Andreev level $\pm \widetilde{E}_A$, 
and the Wilson raton $R$ 
in the large SC gap limit $\Delta_\mathrm{SC}\to \infty$, using the NRG, 
and have deduced the transport properties at $T=0$.

The Bogoliubov particles are strongly  renormalized 
in the Kondo regime 
while the renormalization is a minor  effect 
in the local-Cooper-pairing regime which corresponds 
to the {\it frozen-impurity\/} fixed point of the NRG.
Near the crossover between the two regimes  a pair of 
 renormalized Andreev levels $\pm \widetilde{E}_A$ 
approach the Fermi level, and 
the conductance 
between the dot and normal lead has a peak.
The Josephson current between 
the two SC leads is suppressed significantly in  
the Kondo regime.

We have also presented the spectral function calculated with the NRG. 
The results demonstrate precise features of the original 
Andreev levels, which for the local-Cooper-pairing state are 
broadened by $\Gamma_N$ 
and then renormalized as $U$ increases. 
In the Kondo regime, the pair of 
renormalized Andreev levels overlap to
form a single peak near the Fermi level while at high energies four  
additional peaks are visible. These peaks 
correspond to the excitations 
to the upper and lower  atomic peaks,
which  are defined with respect to the Bogoliubov particles 
and consist of a linear combination of 
the empty and doubly occupied electron states. Thus, 
the broadened bare Andreev peaks and a low energy feature corresponding to a Kondo
resonance can appear within the superconducting gap. For suitable parameters this should
be observable experimentally in the discussed Y-shape geometry.

\begin{acknowledgments}
We thank Yasuhiro Yamada, Rui Sakano, R.\ S.\ Deacon, 
A.\ C.\ Hewson, and N.\ Kawakami for valuable discussions.
A.O.\ is supported by JSPS Grant-in-Aid for
Scientific Research C (No.\ 23540375).
Y.T.\ was supported by Special Postdoctoral Researchers Program of RIKEN.
J.B.\ 
acknowledges financial support from the DFG through the grant BA 4371/1-1.
Numerical computation was partly carried out 
at Yukawa Institute Computer Facility.
\end{acknowledgments}

 \appendix


%

\section{Current in the large gap limit}
\label{sec:appendix_Green}

We provide the expression of the Josephson current 
in the large gap limit in this appendix. 
To this end, we use the imaginary time Green's function, defined by
\begin{align}
\mbox{\boldmath $G$}_{dd}(\tau)
%
\, =& \  -\, 
  \left[ \,
 \begin{matrix}
\langle T_{\tau} \, 
  d_{\uparrow}^{\phantom{\dagger}}(\tau)\, d_{\uparrow}^{\dagger}
 \rangle 
 &
 \langle T_{\tau} \, 
  d_{\uparrow}^{\phantom{\dagger}}(\tau)\, d_{\downarrow}^{\phantom{\dagger}}
 \rangle 
 \cr
 \langle T_{\tau} \, 
   d_{\downarrow}^{\dagger}(\tau) \, d_{\uparrow}^{\dagger}
 \rangle 
 & 
 \langle T_{\tau} \, 
  d_{\downarrow}^{\dagger}(\tau) \, d_{\downarrow}^{\phantom{\dagger}}
 \rangle 
  \end{matrix}
 \, \right]  \;.
\label{eq:G_dd_Nambu}
\end{align}
The Fourier transform of this function can be expressed 
in the form,  
\begin{align}
\Bigl\{ \mbox{\boldmath $G$}_{dd}(i \omega) \Bigr\}^{-1} 
=& \ \,  
i \omega \mbox{\boldmath $1$} 
\,- \, \xi_d\,\bm{\tau}_3 \,+\,i\Gamma_N \,
\mathrm{sgn} \, \omega\,\bm{1}\, 
\nonumber \\
& - \sum_{\nu=L,R} v_{\nu}^2 \, 
\mbox{\boldmath $g$}_{\nu}^{}(i\omega)
\,-\, \mbox{\boldmath $\Sigma$}(i \omega) \;.
\label{eq:Gdd_def}
\end{align}
Here, $i\omega$ is the Matsubara frequency, 
$\mbox{\boldmath $\Sigma$}(i \omega)$ 
is the self energy due to the Coulomb interaction, 
$\mbox{\boldmath $g$}_{\nu}^{}(i \omega)$ is 
the local Green's function at the junction of the lead $\nu$ 
\begin{align}
\mbox{\boldmath $g$}_{\nu}^{}(i\omega)
=
- \, \pi \rho_{\nu}^{\phantom{0}} 
\, 
\frac{ i \omega  \mbox{\boldmath $1$} - 
\mbox{\boldmath $\Delta$}_{\nu}}{
\sqrt{\omega^2 + |\Delta_{\nu}|^2 }
} \;,
\qquad 
\bm{\Delta}_{\nu} \equiv   
\left[ \,
 \begin{matrix}
 0 &  \Delta_{\nu} \cr
 \Delta_{\nu}^* & 0
 \end{matrix}
 \, \right]  ,
\label{eq:G_lead_2}
\end{align}
 $\bm{1}$ and $\bm{\tau}_3$ are the unit 
and Pauli matrices, respectively.

The Josephson current from the dot to the SC lead for $\nu=L,R$ 
can be expressed in terms of the Green's function 
\begin{align}
\langle J_{\nu} \,\rangle 
 =&  \ \frac{e}{\hbar}\, \frac{i  v_{\nu}^2}{\beta} 
\nonumber \\
& \times   
\sum_{\omega_n}
\mbox{Tr}\Bigl[\,
\left\{
\mbox{\boldmath $g$}_{\nu}^{\phantom{0}}(i\omega_n)\,
\mbox{\boldmath $\tau$}_3^{\phantom{0}}
-
\mbox{\boldmath $\tau$}_3^{\phantom{0}}\,
\mbox{\boldmath $g$}_{\nu}^{\phantom{0}}(i\omega_n)
\right\} \, 
\mbox{\boldmath $G$}_{dd}^{\phantom{0}}(i\omega_n)
\,\Bigr] .
\label{eq:J_Joseph_formula}
\end{align}

In the limit of  $|\Delta_{\nu}| \to \infty$,
the lead Green's function becomes a constant  
$\mbox{\boldmath $g$}_{\nu}^{}
\to
\, \pi \rho_{\nu}^{\phantom{0}} 
\, 
\bm{\Delta}_{\nu}/|\Delta_{\nu}|$ as the 
retardation effects caused by the $\omega$ dependence 
in Eq.\ \eqref{eq:G_lead_2} 
are suppressed. Then, Eq.\ \eqref{eq:J_Joseph_formula}
can be rewritten in the form,
\begin{align}
\langle J_{\nu} \rangle  
 \ \to&  \  
\frac{e}{\hbar}\,4\, \Gamma_{\nu}^{\phantom{0}} 
\frac{ 
e^{i\theta_{\nu}} \langle d_\downarrow^{} d_\uparrow^{} \rangle 
-e^{-i\theta_{\nu}} \langle d_\uparrow^\dagger d_\downarrow^\dagger  \rangle 
}{2i}
 \\
 = & \ 
\frac{e}{\hbar}\,4\, \Gamma_{\nu}^{\phantom{0}}
 \left|\langle d_\downarrow d_\uparrow \rangle\right|
 \,\sin\left(\theta_{\nu} -\theta_{d}^{} \right) \;.
\end{align}
Note that 
$\langle d_\downarrow d_\uparrow \rangle 
=\left|\langle d_\downarrow d_\uparrow \rangle\right| 
\,e^{i\theta_{d}^{}}$, 
 in the limit of large gap as   
the phase of  $\langle d_\downarrow d_\uparrow \rangle$ coincides 
with that of the local SC gap 
$\Delta_d 
= |\Delta_d|\,e^{i\theta_{d}^{}}$ 
as shown in Eq.\ \eqref{eq:K_sDQD}.
%
%
%
The current conservation 
$\langle J_{R} \rangle + \langle J_{L} \rangle =0$ 
can be confirmed explicitly through the identity 
\begin{align}
 \Gamma_R^{\phantom{0}}
\,\sin\left(\theta_{R}^{} -\theta_{d}^{} \right)
- \Gamma_L^{\phantom{0}}
\,\sin\left(\theta_{d}^{}-\theta_{L}^{}  \right)
\ = \,0 \;,
\label{eq:conserve}
\end{align}
which follows 
from the definition of $\theta_d$ given in Eq.\ \eqref{eq:Delta_d}. 
%
Using Eq.\ \eqref{eq:conserve}, 
the current can be expressed in the form 
\begin{align}
\langle J \rangle 
\,=\,& 
\frac{e}{\hbar} \,
4 \Gamma_R^{\phantom{0}}
\Gamma_L^{\phantom{0}}\frac{
 \left|\langle d_\downarrow d_\uparrow
 \rangle\right|}
{\left|\Delta_d\right|}
\,\sin \phi \;.
\label{eq:y_junction_current}
\end{align}
This can be rewritten further, 
in terms of the phase shift $\delta$ and 
the Bogoliubov angle $\Theta_B$,  
as shown in Eq.\ \eqref{eq:JosephsonCurrentY-junction}.


\end{document}